\documentclass[a4paper,11pt]{article}
\pdfoutput=1
\usepackage{color}
\usepackage{jheppub}
\usepackage[T1]{fontenc}
\usepackage{amssymb}
\usepackage{graphicx}
\usepackage{amsmath}
\usepackage{hyperref}
\usepackage{tensor}
\usepackage{extarrows}
\usepackage{wrapfig,boxedminipage,setspace,subfigure,epsfig}
\usepackage{tikz}
\usetikzlibrary{decorations.pathmorphing,calc}
\usepackage{epstopdf}
\newcommand{\td}{\text{d}}
\newcommand{\el}{\ell_{\text{AdS}}}

\begin{document}
\title{\boldmath Comparison of holographic and field theoretic complexities by time dependent thermofield double states}
\author[a]{Run-Qiu Yang,}
\author[b]{Chao Niu,}
\author[c,d]{Cheng-Yong Zhang}
\author[b]{Keun-Young Kim,}

\emailAdd{aqiu@kias.re.kr}
\emailAdd{chaoniu09@gmail.com}
\emailAdd{zhangcy0710@pku.edu.cn}
\emailAdd{fortoe@gist.ac.kr}

\affiliation[a]{Quantum Universe Center, Korea Institute for Advanced Study, Seoul 130-722, Korea}
\affiliation[b]{ School of Physics and Chemistry, Gwangju Institute of Science and Technology,
Gwangju 61005, Korea
}
\affiliation[c]{Department of Physics and State Key Laboratory of Nuclear Physics and Technology,
Peking University, 5 Yiheyuan Road, Beijing 100871, China}
\affiliation[d]{Center for High Energy Physics, Peking University, 5 Yiheyuan Road, Beijing 100871, China}

\abstract{
We compute the time-dependent complexity of the thermofield double states by four different proposals:  two holographic proposals based on the ``complexity-action'' (CA) conjecture and ``complexity-volume'' (CV) conjecture, and two quantum field theoretic proposals based on the Fubini-Study metric (FS) and Finsler geometry (FG).
We find that four different proposals yield both similarities and differences, which will be useful to deepen our understanding on the complexity and sharpen its definition.
In particular, at early time the complexity linearly increase in the CV and FG proposals, linearly decreases in the FS proposal, and does not change in the CA proposal. In the late time limit, the CA, CV and FG proposals all show that the  growth rate is $2E/(\pi\hbar)$ saturating the Lloyd's bound, while the FS proposal shows the growth rate is zero. It seems that the holographic CV conjecture and the field theoretic FG method are more correlated.

}
\maketitle
%\tableofcontents
\flushbottom

%%%%%%%%%%%%%%%%%%%%%%%%%%%%%%%%%%%%%%
\noindent

\section{Introduction}
The applications of the holographic principle on the study of black holes have lead to many new surprising discoveries. In particular,  it has been shown that the quantum information may play important roles in the quantum gravity. One important  discovery is the connection between entanglement and geometry~\cite{VanRaamsdonk:2010pw,Ryu:2006bv,Maldacena:2013xja,Faulkner2014}. It inspired a viewpoint that a black hole might be highly entangled with a system that was effectively infinitely far away. This viewpoint lead Juan Maldacena and Leonard Susskind to propose a very interesting conjecture named ``ER=EPR''~\cite{Maldacena:2013xja} when they considered the wormhole created by an Einstein-Rosen (ER) bridge~\cite{PhysRev.48.73} and a pair of maximally entangled black holes. Here EPR refers to quantum entanglement (Einstein-Podolsky-Rosen paradox). To understand how much difficult sending a signal through the ERB from one side to the other, a new information-theoretic quantity named ``complexity'' was imported into the holographic duality and quantum gravity~\cite{Susskind:2014rva,Susskind:2014moa}.  Basically, the complexity describes how many fundamental gates or operators are required when we try to obtain a target state from a reference state.

In order to construct holographic models to describe the complexity, let us consider the following thermofield double (TFD) state
\begin{equation}\label{TFD1}
  |\text{TFD}\rangle:=Z^{-1/2}\sum_\alpha\exp[-E_\alpha/(2T)]|E_\alpha\rangle_L |E_\alpha\rangle_R \,.
\end{equation}
The states $|E_\alpha\rangle_L$ and $|E_\alpha\rangle_R$ are defined in the two copy CFTs and $T$ is the temperature of these two CFTs. This state is conjectured to be approximately dual to an eternal AdS black hole with the Hawking temperature which is the same as $T$ of CFTs~\cite{Maldacena:2001kr}. {With the Hamiltonians $H_L$ and $H_R$ at the left and right dual CFTs, respectively, the time evolution of a TFD state
\begin{equation}\label{timesate1}
  |\psi(t_L,t_R)\rangle:=e^{-i(t_LH_L+t_RH_R)}|\text{TFD}\rangle\,,
\end{equation}
can be characterized by the codimension-two surface of $t=t_L$ and $t=t_R$ at the two boundaries of the AdS black hole~\cite{Maldacena:2001kr,Brown:2015lvg}.} Two different conjectures were proposed to compute the complexity of $|\psi(t_L,t_R)\rangle$ holographically:\footnote{There are also other holographic proposals for complexity, see Refs.~\cite{Alishahiha:2015rta,Ben-Ami:2016qex,Couch:2016exn} for examples. } the CV(complexity=volume) conjecture~\cite{Susskind:2014rva,Stanford:2014jda,Alishahiha:2015rta} and the CA(complexity= action) conjecture~\cite{Brown:2015bva,Brown:2015lvg}.

The CV conjecture states that the complexity of {$|\psi(t_L,t_R)\rangle$} is proportional to the maximal volume of the space-like codimension-one surface which connects the codimension-two time slices denoted by $t_L$ and $t_R$ at  two AdS boundaries, i.e.
\begin{equation}\label{CV}
  \mathcal{C}_V=\max_{\partial \Sigma=t_L\cup t_R}\left[\frac{V(\Sigma)}{G_N \ell}\right] \,,
\end{equation}
where $G_N$ is the Newton's constant. $\Sigma$ is a possible space-like codimension-one surface which connects $t_L$ and $t_R$. $\ell$ is a length scale associated with the bulk geometry such as the horizon radius or AdS radius and so on.
%In this paper, we will choose that $\ell=\el$, where $\el$ is the AdS radius.
However, there is an ambiguity coming from the choice of a length scale $\ell$. This unsatisfactory feature motivated the second conjecture: the CA conjecture \cite{Brown:2015bva,Brown:2015lvg}, which says the complexity of {$|\psi(t_L,t_R)\rangle$} is dual to the action in the Wheeler-DeWitt (WDW) patch associated with $t_L$ and $t_R$, i.e.
\begin{equation}\label{CA}
  \mathcal{C}_A=\frac{I_{\text{WDW}}}{\pi\hbar}.
\end{equation}
The WDW patch associated with $t_L$ and $t_R$ is the collection of all space-like surface connecting $t_L$ and $t_R$ with the null sheets coming from $t_L$ and $t_R$. More precisely it is the domain of dependence of any space-like surface connecting $t_L$ and $t_R$.
%This conjecture has some advantages compared with the CV conjecture. For example, it has no free parameter and can satisfy Lloyd's complexity growth bound in very general cases \cite{Lloyd2000,Cai:2016xho,Yang:2016awy,Cai:2017sjv}. However, the CA conjecture has its own obstacle in computing the action: it involves null boundaries and joint terms. Recently, this problem has been overcome by carefully analyzing the boundary term in null boundary \cite{Parattu:2015gga,Lehner:2016vdi}.

Recently, two different methods were proposed by Refs.~\cite{Chapman:2017rqy,Yang:2017nfn} to define the complexity in quantum field theory.\footnote{Refs.~\cite{Caputa:2017urj,Caputa:2017yrh} also give a definition for the complexity in conformal field theory by the Liouville action. It is also interesting to see what the results are if this method is applied to the time-dependent TFD states. } The method in Ref.~\cite{Chapman:2017rqy}, which we will call the FS method in this paper, is based on the Fubini-Study metric and defined the complexity of two states to be the length of the geodesic under the Fubini-Study metric. The method in Ref.~\cite{Yang:2017nfn},  which we will call the FG method in this paper, first defined the complexity for operators by the Finsler geometry\footnote{ The Finsler geometry was first introduced to investigate the computational complexity by Refs.~\cite{Nielsen1133,Nielsen:2006:GAQ:2011686.2011688,Dowling:2008:GQC:2016985.2016986} and recently drew attention again in Ref.~\cite{Jefferson:2017sdb}.} and then used this complexity to define the complexity between states.
%Refs.~\cite{Chapman:2017rqy,Yang:2017nfn} checked these two different proposals in some models and found some very interesting results. Especially,
As what we will present later, these two methods are very suitable to compute the complexity between two different TFD states.

Taking the Eqs.~\eqref{CV}, \eqref{CA} and two quantum field theory proposals in Refs.~\cite{Yang:2017nfn,Chapman:2017rqy} into account, we have at least four different methods to compute the complexity between two TFD states.
The goal of this paper is to compute the complexity in four different methods and see their similarities and differences. It may give us some information to judge which are appropriate methods to compute the complexity among four methods. It may also shed light on a possible connection between two holographic conjectures and two quantum field theory proposals.

The paper is organized as follows. In sec.~\ref{modefCVCA} we introduce several concepts for complexity in literature and clarify some subtle issues in defining the complexity regarding the reference state and divergence in holographic complexity.  In sec.~\ref{CACV} we use the modified CA and CV conjectures introduce in sec.~\ref{modefCVCA} to compute complexity between the time dependent TFD state and their corresponding zero temperature vacuum state.
In sec.~\ref{fromQTF},  after presenting  how to construct a time-dependent TFD state for free field theory explicitly, we use two different quantum field theoretic proposals to compute the complexity between the time-dependent TFD state and the vacuum. Our computations by four different methods are summarized and compared in sec.~\ref{summ}.

%\kyr{We had a citation request by Mohammad. please cite the following paper \cite{Qaemmaqami:2017lzs} somewhere.}

\paragraph{Note added}
While this work was being finished, Ref.~\cite{Carmi:2017jqz} appeared which also studied the complexity growth rate.
Our results in sec.~\ref{CACV} have some overlap with Ref.~\cite{Carmi:2017jqz}.

\section{Holographic complexity potential}\label{modefCVCA}
Before we compute the holographic time-dependent complexity, let us first make some comments on the CV and CA conjectures and present our modified versions of them.

Although these two conjectures satisfy important requirements on the complexity such as the Lloyd's bound~\cite{Lloyd2000,Cai:2016xho,Yang:2016awy,Cai:2017sjv}, it seems that there are a few subtle issues to be clarified. First, the complexity computed by Eq.\eqref{CV} or Eq.~\eqref{CA} is infinite. Second, there is an ambiguity for the reference state.
One may assume that there is an unknown ``favorite'' reference state which is not found yet\footnote{ Because the complexity from any state to itself is zero, if this reference state is any state dual to an asymptotic AdS black hole then we can find Eq.~\eqref{CV} or \eqref{CA} should be zero at this black hole. However, Ref.~\cite{Carmi:2016wjl} has proven that Eqs.~\eqref{CV} or \eqref{CA} are divergent for all asymptotically  AdS black holes, so this reference state is not dual to any AdS black hole.  }, and the divergence of the complexity computed by the CV and CA conjectures show some intrinsic properties of CFT similarly to the case of the entanglement entropy~\cite{Carmi:2016wjl}.
Even if we accept this unknown ``favorite'' reference state exists, it seems that the original CA conjecture has two more issues. First, in principle, the complexity between two states should be non-negative but the value computed by Eq.~\eqref{CA} can be negative.  Second, the dynamics will be invariant if we add a constant term into the action so the complexity should also be invariant after adding a constant term into the gravity action. However, the original CA conjecture does not satisfy this property.

To resolve this issue a ``modified'' complexity was proposed in Ref.~\cite{Yang:2017nfn}. It suggests that the original CV and CA conjectures \eqref{CV} and  \eqref{CA} give a kind of ``complexity potential'' rather than the complexity between any two states. When we restrict our considerations to the TFD  states, the leading order of complexity between two TFD states $|\text{TFD}_1\rangle$ and $|\text{TFD}_2\rangle$ is given by the following formulas in the CV or CA conjectures\footnote{Formally, Eq.~\eqref{newcvca} is similar to the ``complexity of formation'' proposed by Ref.~\cite{Chapman:2016hwi}. However, they are not always the same. A detailed discussion about it can be found in Ref.~\cite{Yang:2017nfn}. }
\begin{equation}\label{newcvca}
\begin{split}
  &\mathcal{C}_V(|\text{TFD}_2\rangle,|\text{TFD}_1\rangle)=|\mathcal{C}^{(1)}_{V}-\mathcal{C}^{(2)}_{V}|\,, \\
  &\mathcal{C}_A(|\text{TFD}_2\rangle,|\text{TFD}_1\rangle)=|\mathcal{C}^{(1)}_{A}-\mathcal{C}^{(2)}_{A}|\,.
  \end{split}
\end{equation}
Here $\{\mathcal{C}^{(1)}_{V}, \mathcal{C}^{(2)}_{V}\}$ and  $\{\mathcal{C}^{(1)}_{A}, \mathcal{C}^{(2)}_{A}\}$ are computed by Eqs.~\eqref{CV} or \eqref{CA}. This modification does not lose any important physical properties of the original version and seems simpler because it does not need to refer to an unknown reference state\footnote{In principle, a well-defined complexity should satisfy that $\mathcal{C}(|\text{TFD}_2\rangle,|\text{TFD}_1\rangle)=0$ if and only if $|\text{TFD}_2\rangle=|\text{TFD}_1\rangle$. However, it may be possible that the modified version in Eq.~\eqref{newcvca} vanish even with two different states. This can appear if the system has multiple different solutions  for given physical conditions, for example, in the cases which contain phase transitions. Another possibility stems from the  the fact that the TFD states is only approximately dual to the eternal AdS black holes so we can expect the original CA and CV conjectures in Eq.~\eqref{CV} and \eqref{CA} may lose some subleading contributions.}.  This modified version has the following basic properties:
%
%(1) $\mathcal{C}(|\text{TFD}_2\rangle,|\text{TFD}_1\rangle)=0$ if and only if $|\text{TFD}_2\rangle=|\text{TFD}_1\rangle$;\\
\begin{itemize}
\item Triangle inequality: $\mathcal{C}(|\text{TFD}_2\rangle,|\text{TFD}_0\rangle)+\mathcal{C}(|\text{TFD}_0\rangle,|\text{TFD}_1\rangle)\geq\mathcal{C}(|\text{TFD}_2\rangle,|\text{TFD}_1\rangle$ for any state $|\text{TFD}_0\rangle$
\item Reversibility: $\mathcal{C}(|\text{TFD}_2\rangle,|\text{TFD}_1\rangle)=\mathcal{C}(|\text{TFD}_1\rangle,|\text{TFD}_2\rangle)$
\end{itemize}
In particular, when Eq.~\eqref{newcvca} is applied to the TFD states, the results in Eq.~\eqref{newcvca} agree to the results obtained in a quantum field theory approach in Ref.~\cite{Yang:2017nfn} and also agree to the results computed by the method proposed in Ref.~\cite{Chapman:2017rqy} (which will be presented in subsection~\ref{FSmetric0}).

The formula \eqref{newcvca} can be understood more geometrically.
%Let us assume $\mathcal{H}$ is the set of all the states in holographic duality.
Let us assume a space where all states can be parameterized by $x^a$, which may be  the temperature, total charge, mass or
any other quantity describing different time slices at two boundaries.
%or any other parameter to describe bulk geometry and boundary time slices.
Suppose that there is a curve $l: x^a=x^a(\lambda)$ which satisfies $|\text{TFD}(x^a(\lambda_1))\rangle=|\text{TFD}_1\rangle$ and $|\text{TFD}(x^a(\lambda_2))\rangle=|\text{TFD}_2\rangle$ with $\lambda_1\leq\lambda_2$. For any given curve $l$, we can use  Eqs.~\eqref{CV} or \eqref{CA} to compute the $\mathcal{C}_V(\lambda)$ and $\mathcal{C}_A(\lambda)$, which are the functions of $\lambda$ and depend on the choice of the curve $l$. Then the complexity between $|\text{TFD}_1\rangle$ and $|\text{TFD}_2\rangle$ is given by\footnote{We assume that the curve is regular, which means that $0\leq|\partial x^a/\partial\lambda|<\infty$.}
\begin{equation}\label{newcvca2v}
  \mathcal{C}_V(|\text{TFD}_2\rangle,|\text{TFD}_1\rangle)=\min\left\{\int_{\lambda_1}^{\lambda_2}\left.\left|\frac{\td \mathcal{C}_V(\lambda)}{\td \lambda} \right|\td \lambda~\right|~~\forall~l:x^a=x^a(\lambda)\right\}\,,
\end{equation}
and
\begin{equation}\label{newcvca2a}
  \mathcal{C}_A(|\text{TFD}_2\rangle,|\text{TFD}_1\rangle)=\min\left\{\int_{\lambda_1}^{\lambda_2}\left.\left|\frac{\td \mathcal{C}_A(\lambda)}{\td \lambda} \right|\td \lambda~\right|~~\forall~l:x^a=x^a(\lambda)\right\}\,.
\end{equation}
These two equations in fact give the Finsler structures such that for any tangent vector  $T^a=(\partial/\partial\lambda)^a=\td x^a/\td\lambda$
\begin{equation}\label{finsler}
  F(x^a,T^a):=\left|(\textbf{\td}\mathcal{C}_X)_a T^a\right|\,,
\end{equation}
where $X=V$ or $A$ and $\textbf{\td}$ is the exterior differential operator in the parameter space spanned by $x^a$. The integrations in Eqs.~\eqref{newcvca2v} and \eqref{newcvca2a}  give a holographic version of ``complexity geometry'', which is similar to the FG and FS methods in Refs.~\cite{Chapman:2017rqy,Chapman:2016hwi,Yang:2017nfn}. Since the absolute value appearing in these two formulas is not convenient, we introduce a positive infinitesimal value $\epsilon$ and arbitrary functions $\omega_{ab}$  so that Eqs.~\eqref{newcvca2v} and \eqref{newcvca2a} can be written as the following form
\begin{equation}\label{newcvca2X}
  \mathcal{C}_X(|\text{TFD}_2\rangle,|\text{TFD}_1\rangle)=\lim_{\epsilon\rightarrow0^+}\min\left\{\int_{\lambda_1}^{\lambda_2}\sqrt{\left(\frac{\partial\mathcal{C}_X}{\partial x^a}\frac{\td x^a}{\td\lambda}\right)^2+\epsilon^2\omega_{ab}\frac{\td x^a}{\td\lambda}\frac{\td x^b}{\td\lambda}}\td\lambda\right\}\,.
\end{equation}
It is assumed that the limit is well defined and independent of the choices of auxiliary functions $\omega_{ab}$. This means that we have the following ``holographic complexity metric'' defined in a parameter space spanned by $x^a$,
\begin{equation}\label{holocomg1}
  \td s_X^2:=\left[\frac{\partial\mathcal{C}_X}{\partial x^a}\frac{\partial\mathcal{C}_X}{\partial x^b}+\epsilon^2\omega_{ab}\right]\td x^a\td x^b,
\end{equation}
Since the metric defined in Eq.~\eqref{holocomg1} is a tensor the complexity is diffeomorphism invariant under the reparameterizations on $x^a$. The minimal values of Eq.~\eqref{newcvca2v} then is the lengths of geodesics given by metric \eqref{holocomg1} if $\mathcal{C}_V(x^a)$ and $\mathcal{C}_A(x^a)$ are $C^2$ functions of $x^a$.

%This means the Eq.~\eqref{newcvca2X} will be not the same as the Eqs.~\eqref{newcvca} in general cases.
To show that Eqs.~\eqref{newcvca2v} and \eqref{newcvca2a} are equivalent to Eq.~\eqref{newcvca} note that
\begin{equation}\label{notemetrics}
  \int_{\lambda_1}^{\lambda_2}\td\lambda\left|\frac{\td \mathcal{C}_{X}}{\td\lambda}\right|=\int_l|\td\mathcal{C}_X|\geq\left|\int_l\td\mathcal{C}_X\right|=|\mathcal{C}_X^{(1)}-\mathcal{C}_X^{(2)}|\,.
\end{equation}
The equality can be achieved if there is a curve with its tangent $T^a$ satisfying,
\begin{equation}\label{conditionTs}
  (\textbf{\td}\mathcal{C}_X)_aT^a\leq0, \ \  \forall \lambda\in(\lambda_1,\lambda_2), \quad \text{or} \quad (\textbf{\td}\mathcal{C}_X)_aT^a\geq0, \ \ \forall \lambda\in(\lambda_1,\lambda_2)\,.
\end{equation}
For the case that parameter space has trivial topology and its dimensional is larger than 1, such a curve always can be found, so Eq.~\eqref{notemetrics} shows,
\begin{equation}\label{newcvca3a}
  \mathcal{C}_X(|\text{TFD}_1\rangle,|\text{TFD}_2\rangle)=|\mathcal{C}_{X}^{(1)}-\mathcal{C}_{X}^{(2)}|\,,
\end{equation}
which is Eq.~\eqref{newcvca}. However, for the cases that parameter space is one dimensional or has non-trivial topology, the condition~\eqref{conditionTs} may not be achieved. The complexity modified definitions~\eqref{newcvca2v} and \eqref{newcvca2a} may different from ones in Eq.~\eqref{newcvca3a}. In these cases, one have to use definitions~\eqref{newcvca2v} and \eqref{newcvca2a} to compute the complexity in holography.
%However, there are still a few case that cannot field any curve which can satisfy Eq.~\eqref{conditionTs}. In these cases, we have to solve Eqs.~\eqref{newcvca2v} and \eqref{newcvca2a} or use holographic complexity metric~\eqref{holocomg1} to find the length of geodesic. In following section when we compute the complexity between a TFD state and its corresponding zero temperature vacuum state, we will see that the condition \eqref{conditionTs} is always satisfied expect for BTZ black holes in CA conjecture.

Let us make a summary regarding several concepts for the complexity introduced in literature and this paper. If we can compute the ``complexity potential'' $\mathcal{C}_V(x^a)$ and $\mathcal{C}_A(x^a)$ by Eqs.~\eqref{CV} and \eqref{CA}, which is the original proposals, then we can obtain the ``modified complexity'' which is the complexity between two states through Eq.~\eqref{newcvca3a}.
The holographic complexity potential has an additional freedom: if we add any term independent of $x^a$, the modified complexity will be invariant. This freedom gives us a possibility to introduce suitable subtraction terms to renormalize the divergent holographic complexity potential defined by Eqs.~\eqref{CV} and \eqref{CA}. This is the foundation of the ``regularized holographic complexity'' proposed by Ref.~\cite{Kim:2017lrw}, which we will call ``renormalized complexity potential'' in this paper.
%Ref.~\cite{Kim:2017lrw} proposed an method to renormalized the holographic complexity potential by adding some suitable counterterm. For planar symmetric AdS black holes and flat time slices, these counterterm are independent of the bulk geometry and so independent of $s^a$. This means that we can use renormalized holographic complexity potential to compute the complexity between two TFD states.

\section{Time dependent complexity of the TFD states: holographic approach}\label{CACV}
In this section, we will use the modified CV and CA conjectures to compute the complexity between the time-dependent TFD states and their corresponding vacuum states. There are three parameters in the holographic duals: the temperature of the bulk black hole, the  time of the left boundary and the time of the right boundary slice i.e. $\{T, t_L, t_R\}$. We first have to compute the complexity potential $\mathcal{C}_V(T,t_L, t_R)$ and $\mathcal{C}_A(T,t_L, t_R)$ which are divergent. To deal with this divergence, Ref.~\cite{Kim:2017lrw} proposed a method to renormalize them by adding some counterterms. In the planner symmetry AdS black holes,  the general counterterms have been found and are independent of the values of $T, t_L$ and $t_R$. Therefore, we can use the renormalized complexity potential for the holographic complexity potential and finally find the modified complexity between two TFD states. In the following subsections, we will perform these procedures in both the CA and CV conjectures.

Since we focus on the TFD states which are dual to the planar symmetric Schwarzschild AdS black holes, the problems can be simplified. Thanks to the time translation symmetry, the systems only depend on $t_L+t_R$ so  we are left with only two independent parameters $\{T, t_L+t_R\}$.  %Secondly, as the system has scaling symmetry $T\rightarrow \alpha T, \Sigma_{d-1}\rightarrow \alpha^{1-d}$, we can always fix

\subsection{CA conjecture}
\subsubsection{Total action with the boundary and joint terms}\label{CAconj}
The central issue in the CA conjecture is the computation of the on-shell action. Because the null boundary and joint terms are involved in this computation, it was a subtle problem at the time when this conjecture was proposed. However, after several careful analysis on the action with null boundaries~\cite{Parattu:2015gga,Lehner:2016vdi,Hopfmuller:2016scf,Jubb:2016qzt}, the action in general relativity with suitable boundary and joint terms turned out to be the following form,
\begin{equation}\label{actionull}
  I=\frac1{16\pi}\int_{\mathcal{M}}\td^{d+1}x\sqrt{-g}\left[R+\frac{d(d-1)}{\el^2}\right]+\sum_{i}I_{B_i}+\sum_jI_{\mathcal{N}_j}+\sum I_{\mathcal{J}}+\sum I_{\mathcal{J}'}\,.
\end{equation}
Here $R$ is the scalar curvature in the bulk, $\el$ is the AdS radius, $I_{B_i}$ is the Gibbons-Hawking-York (GHY) boundary term for the non-null boundary fragments  $B_i$, $I_{\mathcal{N}_j}$ is the boundary term for the null boundary fragments $\mathcal{N}_j$, $I_{\mathcal{J}}$ is the joint term defined on the joint of two non-null boundaries and $I_{\mathcal{J}'}$ is the joint term defined  on the joint intersected by the null boundaries and others.

The first result for the null boundary term was proposed by Ref.~\cite{Parattu:2015gga} and then represented by Refs.~\cite{Lehner:2016vdi,Hopfmuller:2016scf,Jubb:2016qzt}. Refs.~\cite{Parattu:2015gga,Jubb:2016qzt} showed that the suitable null boundary term should be
\begin{equation}\label{nullbd1}
  I_{\mathcal{N}_j}=I_{\mathcal{N}_j}^{(1)}:=-\frac{\text{sign}(\mathcal{N}_j)}{8\pi}\int_{\mathcal{N}_j}\td\lambda\td^{d-2}x\sqrt{|\sigma|}(\kappa+\Theta)\,,
\end{equation}
where $\lambda$ is the integral curve of the normal vector $k^\mu$ (future directed) for the null boundary, i.e., $k^\mu=(\partial/\partial\lambda)^\mu$. $\kappa$ is the ``surface gravity'' of the null surface corresponding to $k^\mu$ and satisfies $k^\mu\nabla_\mu k^\nu=\kappa k^\nu$. $\text{sign}(\mathcal{N}_j)$ is +1 only when it lies on the future boundary of $\mathcal{M}$, otherwise, $\text{sign}(\mathcal{N}_j)$ is $-1$. $\sigma$ is the determinant of the induced metric at the transverse co-dimensional 2 surface orthogonal to $k^\mu$.   $\Theta$ is the expansion of the null boundary measured by $\lambda$ and satisfies $\Theta=(\sqrt{|\sigma|})^{-1}k^\mu\partial_\mu\sqrt{|\sigma|}$. Considering the fact that $\Theta$ itself vanishes during the variation, Refs.~\cite{Lehner:2016vdi,Hopfmuller:2016scf} proposed the ``minimal null boundary term'' by dropping the expansion term in Eq.~\eqref{nullbd1}, which reads
\begin{equation}\label{nullbd2}
  I_{\mathcal{N}_j}=I_{\mathcal{N}_j}^{(2)}:=-\frac{\text{sign}(\mathcal{N}_j)}{8\pi}\int_{\mathcal{N}_j}\td\lambda\td^{d-2}x\sqrt{|\sigma|}\kappa\,.
\end{equation}
%
%Though the expansion term in $I_{\mathcal{N}}^{(1)}$ is not necessary to insure the variational problem to be well defined, it has special significance.  Ref.~\cite{Parattu:2015gga} has shown that this  null boundary term \eqref{nullbd1} can be obtain from the GHY boundary term using the timelike/spacelike surface to approach to the null limit. Ref.~\cite{Parattu:2016trq,Jubb:2016qzt} has shown that $I_{\mathcal{N}}^{(1)}$ and GHY can be written as an uniform form.
However, the boundary terms \eqref{nullbd1} and \eqref{nullbd2} are both dependent on the choice of $\lambda$ so the re-parameterization of $\lambda$ can lead different values for the null boundary term. To overcome this problem, Refs.~\cite{Lehner:2016vdi,Reynolds:2016rvl} proposed that we should add an additional term into the   boundary term \eqref{nullbd2}
\begin{equation}\label{nullbd4}
  I_{\mathcal{N}_j}=I_{\mathcal{N}_j}^{(3)}:=-\frac{\text{sign}(\mathcal{N}_j)}{8\pi}\int_{\mathcal{N}_j}\td\lambda\td^{d-2}x\sqrt{|\sigma|}(\kappa+L_0)\,,
\end{equation}
where $L_0=\Theta\ln(|\Theta|/\el)$ if  $\mathcal{N}$ lies to the future boundary of $\mathcal{M}$. Otherwise, $L_0=-\Theta\ln (|\Theta|/\el)$. In this paper, we will use this total boundary term.
%The total boundary term in this paper then is,
%%
%\begin{equation}\label{nullbd4}
%  I_{\mathcal{N}}=-\frac{\text{sign}(\mathcal{N}_j)}{8\pi}\int_{\mathcal{N}_j}\td\lambda\td^{d-2}x\sqrt{|\sigma|}(\kappa+\Theta+L_0)\,,
%\end{equation}
%%
%where $c_1$ and $c_0$ and only be 0 or 1. Of cause, only from the viewpoint of variational problem, these two coefficients can be any two real values.
%We will see later that only when $c_0=c_1=1$, can the formation of complexity be positive and its growth rate is bounded and not larger then two times of total energy of black holes.

The joint term $I_{\mathcal{J}}$ was first found by Ref.~\cite{PhysRevD.47.3275} and then confirmed by Refs.~\cite{Lehner:2016vdi,Jubb:2016qzt} again recently by different methods. As the CA conjecture will not have such kind of joints, we will not give the detailed form for this kind of joint terms. The joint term for the case that there is at least one null boundary was first found by Ref.~~\cite{Lehner:2016vdi}, which is in general expressed as
\begin{equation}\label{nulljoint1}
  I_{\mathcal{J}'}=\frac{\text{sign}(\mathcal{J'})}{8\pi}\int_{\mathcal{J'}}\td^{d-1}x\sqrt{\sigma}a\,.
\end{equation}
Here $\sigma$ is the determinant of the induced metric on the joint $\mathcal{J'}$. According to the properties of the intersectional surface, the term $a$ can be computed as
\begin{equation}\label{expressa}
  a=\left\{
  \begin{split}
  &\ln(|n^I k_I|)\,,\\
  &\ln(|k^I\bar{k}_I|/2)\,,
  \end{split}
  \right.
\end{equation}
where $n^I$ is the unit normal vector (outward/future directed) for the non-null intersecting boundary, and $\bar{k}^I$ is the other null normal vector (future directed) for the null intersecting boundary. The value of $\text{sign}(\mathcal{J'})=\pm1$, which can be assigned as follows: ``+1'' appears only when the WDW patch appears in the future/past of the null boundary component and the joint is at the past/future end of the null boundary fragments.
%$a_0$ is any scalar field defined at the joints and invariant under the variation. It has been shown in Ref.~\cite{Lehner:2016vdi} that this freedom comes from that we can redefine the scalar field $\Phi$ that describe the null hypersurface (i.e., with $\Phi=0$). In principle, $a_0$ can be any scalar field define with vanishing variation. This freedom leads an ambiguity in defining the complexity. We will address this problem later on.

\subsubsection{Complexity potential}
%\subsubsection{Time dependence of regularized complexity for AdS$_{d+1}$ black hole}\label{BTZtime}
The metric for the general AdS$_{d+1}$($d\geq2$) black hole is
\begin{equation}\label{metricBTZ}
  \td s^2=-r^2f(r)\td t^2+\frac{\td r^2}{r^2f(r)}+r^2\td\Sigma_{d-1}^2 \,,
\end{equation}
where $\td\Sigma_{d-1}^2=\el^{-2}\sum_{i=1}^{d-1}\td x_i^2$ is the  $(d-1)$ dimensional line element and $\Sigma_{d-1}$ is the volume of the conformal boundary of the AdS black hole.
%, which is given by,
%%
%\begin{equation}\label{sigmaline1}
%  \td\Sigma_{k,d-1}^2=\left\{
%  \begin{split}
%  \td\omega_{d-1}^2=\td\theta^2+\sin^2\theta\td\Omega_{d-2}^2,~~~&\text{for}~k=1\,,\\
%  \el^{-2}\sum_{i=1}^{d-1}\td x_i^2,~~~&\text{for}~k=0\,,\\
%  \td\omega_{d-1}^2=\td\theta^2+\sinh^2\theta\td\Omega_{d-2}^2,~~~&\text{for}~k=1\,.
%  \end{split}
%  \right.
%\end{equation}
%%
%Here $\Omega_{d-2}^2$ is the standard line element of $d-2$ dimensional sphere. We see that $k=1,0$ and $-1$ corresponds to the black hole with spherical, planar and hyperbolic symmetries, respectively.
The function $f(r)$ reads
\begin{equation}\label{BTZfr}
  f(r)=\frac1{\el^2}\left(1-\frac{r_h^d}{r^d}\right)\,.
\end{equation}
The physical total mass (ADM mass) and temperature of this system are
\begin{equation}\label{massAdS}
  M=\frac{r_h^d(d-1)\Sigma_{d-1}}{16\pi\el^2}\,, \qquad T=\frac{r_hd}{4\pi\el^2}\,.
\end{equation}

\begin{figure}
  \centering
  \includegraphics[width=.49\textwidth]{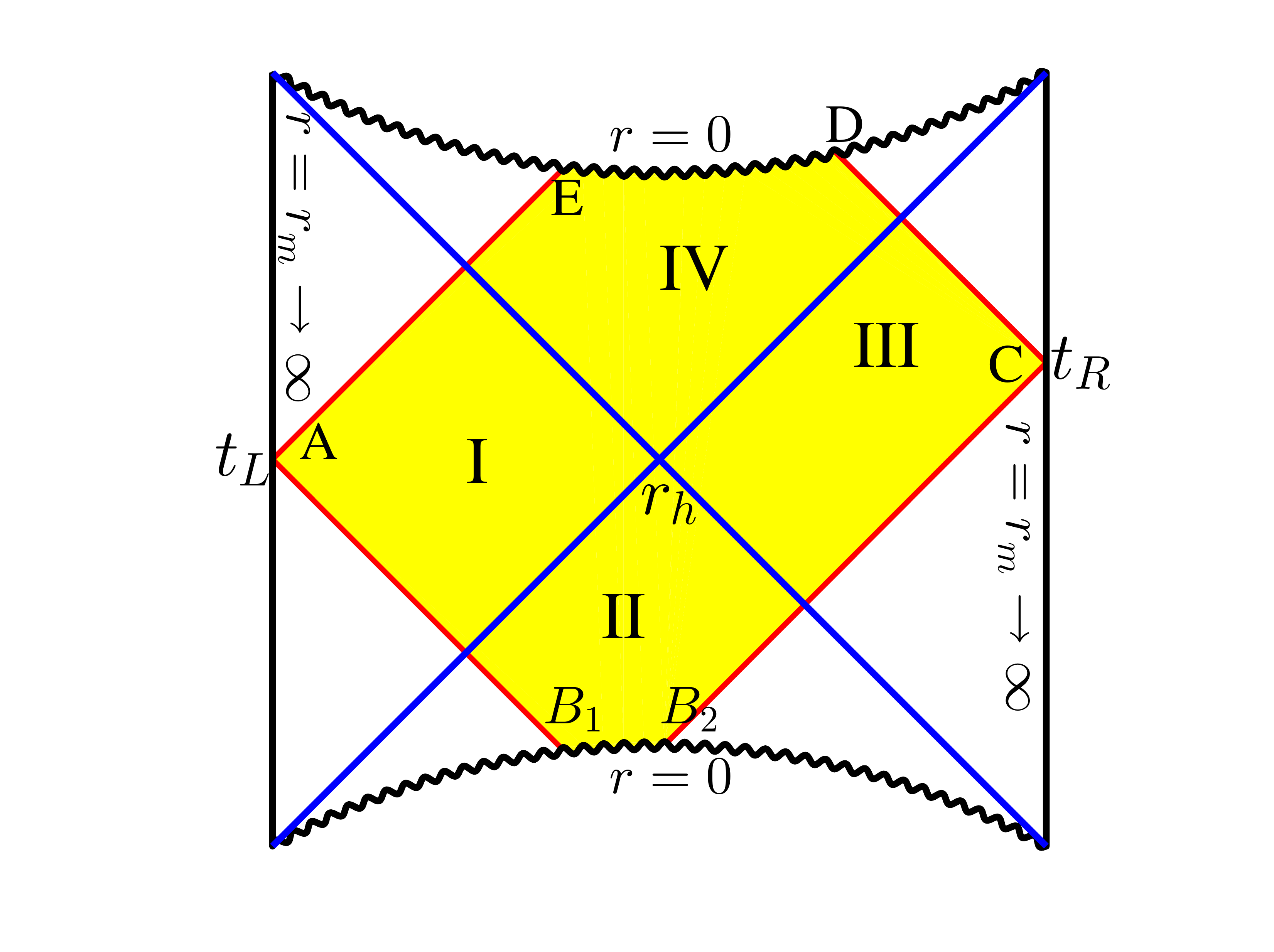}
   \includegraphics[width=.49\textwidth]{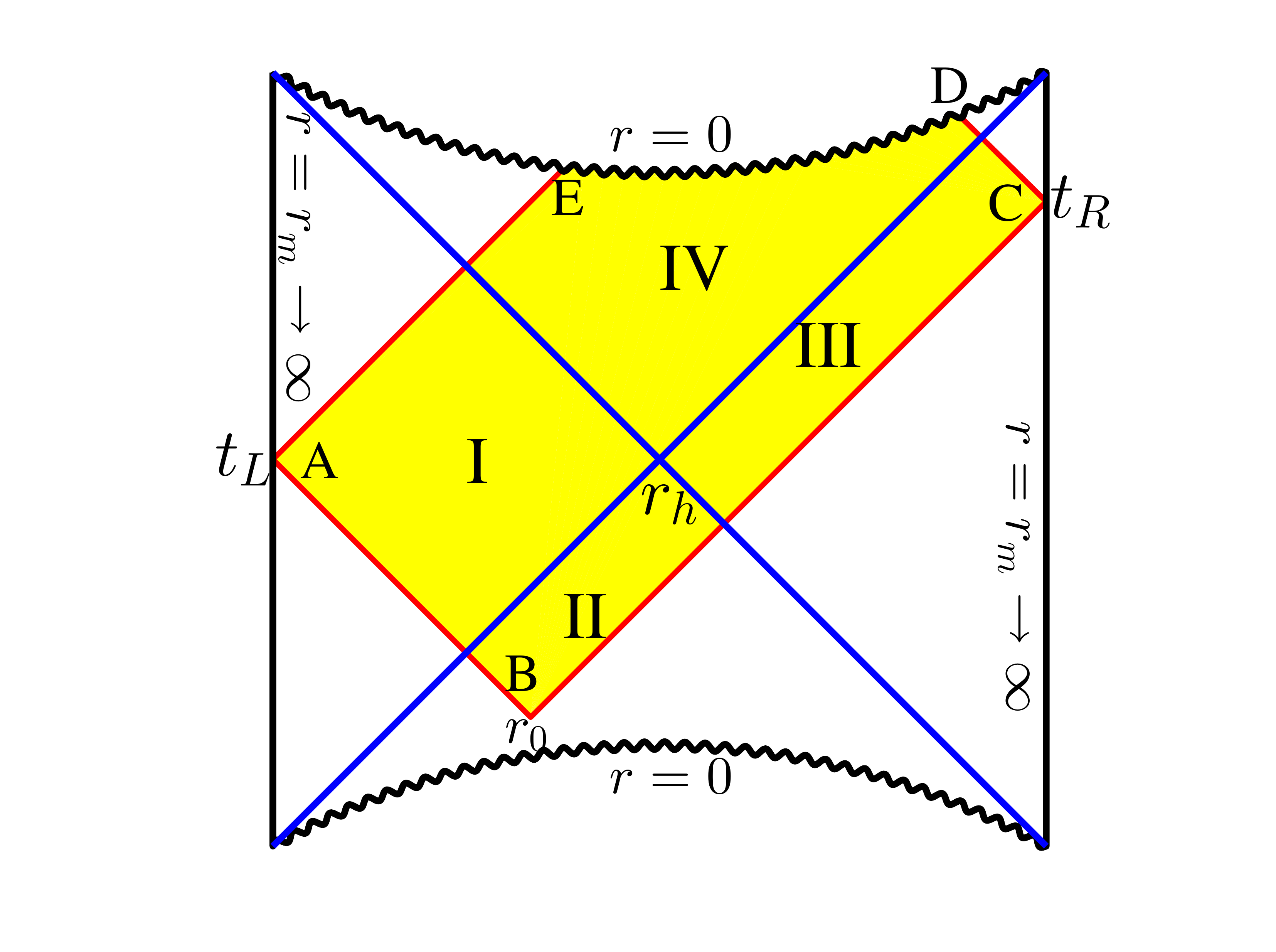}
  \caption{Penrose diagrams and WDW patches for AdS$_{d+1}$ black hole ($d\geq3$) when $|t_R|<\Delta t_c$ (left panel) and $|t_R|>\Delta t_c$ (right panel). In the left panel, the past null sheets will meet the singularity at $B_1$ and $B_2$ respectively. In the right panel, the past null sheets will meet each other at $B$ with $r=r_0\in(0,r_h)$. }\label{SAdS1}
\end{figure}

The Penrose diagram and the WDW patch are shown in Fig.~\ref{SAdS1}.  The time direction of the right boundary is the same as the coordinate time $t$ but the time of the left boundary is opposite to the coordinate time $t$. As the space-time has time translation symmetry the on-shell action only depends on the value of $t_L+t_R$.
By this property we can fix $t_L=0$ and only study how the complexity depends on the value of $t_R$. In addition, thanks to the time reversal symmetry of the black hole we only consdier the case of $t_R>0$.

In Fig.~\ref{SAdS1} we see that there is a critical time $\Delta t_c$ distinguishing the left and right panel. Depending on the relationship between $t_R$ and $\Delta t_c$, there are two different types for the WDW patches. One is the case that $|t_R|<\Delta t_c$ shown in the left panel of Fig.~\ref{SAdS1}, where the two future and past null sheets all meet the singularities. The other one is the case that $|t_R|>\Delta t_c$ shown in the right panel of Fig.~\ref{SAdS1}, where the future directed null sheets coming from the boundaries  will first meet the singularity $r=0$ but the past directed null sheets coming from the boundaries  will meet each other in the inner region of black hole.

Let us introduce the infalling coordinate $v$ and outgoing coordinate $u$ as
\begin{equation}\label{uvcoord1}
  v=t+r^*,~~~u=t-r^*\,,
\end{equation}
where
\begin{equation}\label{AdSrstar}
  r^*(r)=\int^{r}\frac{\td x}{x^2f(x)}\,.
\end{equation}
The null dual normal vector field for the null boundaries $AB$ and $CD$ is $k_\mu=-[(\td t)_\mu+r^{-2}f^{-1}(\td r)_\mu]$ and the null normal vector field for the null boundaries $BC$ and $AE$ is $\bar{k}_\mu=-[(\td t)_\mu-r^{-2}f^{-1}(\td r)_\mu]$.  These two null vector are affinely parameterized. The integration in Eq.~\eqref{AdSrstar} can be expressed in terms of the hypergeometrical function
\begin{equation}\label{solverstar1}
  r^*(r)=\frac{\el^2}{r}\left[_2F_1\left(1,-\frac{1}{d};1-\frac{1}{d}; \frac{r^d}{r_h^d}\right)-1\right]\,, \qquad r<r_h\,,
\end{equation}
and
\begin{equation}\label{solverstar2}
  r^*(r)=\frac{\el^2}{r_h}\left[\cot\frac{\pi}{d}-\frac{r_h}{(r^d-r_h^d)^{1/d}} {_2}F_1\left(\frac1d,\frac1d;1+\frac1d;\frac{r_h^d}{r_h^d-r^d}\right)-1\right]\,, \qquad r>r_h\,.
\end{equation}
The value of $\Delta t_c$  is given by,
\begin{equation}\label{deltatcvalue}
  \Delta t_c:=2[r^*(\infty)-r^*(0)]=\frac{2\el^2}{r_h}\frac{\pi}{d}\cot\frac{\pi}{d}\,.
\end{equation}
To regularize the WDW patch, we assume the AdS boundaries are located at $r=r_m\gg\el$. One can see from  Fig.~~\ref{SAdS1} that when $t_R\leq\Delta t_c$, the WDW patch is the same as the one of $t_R=0$. This means that the corresponding TFD state is the same one for $t_R\in(0,\Delta t_c)$. So we have
\begin{equation}\label{CAt1}
  |\text{TFD}(T,t_R)\rangle= |\text{TFD}(T,0)\rangle,~~\text{if}~|t_R|\leq\Delta t_c\,,
\end{equation}
and
\begin{equation}\label{CAt10}
  \mathcal{C}_{\text{A,ren}}(t_R)=\mathcal{C}_{\text{A,ren},0}\,,
\end{equation}
where $\mathcal{C}_{\text{A,ren}}(t_R)$ is the renormalized complexity potential for given $t_R$ and $\mathcal{C}_{\text{A,ren},0}$ is its value when $t_R=0$. The value of $\mathcal{C}_{\text{A,ren},0}$ has been given by Ref.~\cite{Chapman:2016hwi,Kim:2017lrw},
\begin{equation}\label{CAregs0}
  \mathcal{C}_{\text{A,ren},0}=\frac{d-2}{d-1}\cot\left(\frac{\pi}{d}\right)\frac{M}{\hbar T}\,.
\end{equation}
The TFD state begins to evolve after $|t_R+t_L|>\Delta t_c$. In this case, the two past null sheets will meet each other at the joint $B$ with $r=r_0\in(0,r_h)$. We can obtain the equations for all the null boundaries, which are
\begin{equation}\label{lineAB2}
\begin{split}
  AB&:~r^*_m=t+r^*(r)\,,\\
  CD&:~t_R+r^*_m=t+r^*(r)\,,\\
  BC&:~t_R-r^*_m=t-r^*(r)\,,\\
  AE&:~-r^*_m=t-r^*(r)\,.
  \end{split}
\end{equation}
By the equations for $AB$ and $BC$, we find that the past null sheets will meet each other at $r=r_0$, where $r_0$ is defined by the following equation
\begin{equation}\label{eqforr0}
  r^*(r_0)=r^*_m-\frac{t_R}2\,, \qquad r_0<r_h\,.
\end{equation}
%
%As for general cases one will obtain that $r^*(0)\leq r^*_m$ when $d\geq2$,
Eq.~\eqref{eqforr0} has a solution only when $t_R\geq\Delta t_c$. By using Eqs.~\eqref{solverstar1}, \eqref{solverstar2} and taking $r_m\rightarrow\infty$, we find $r_0$ is given as
\begin{equation}\label{valuer0s1}
  \frac{\el^2}{r_0}\left[_2F_1\left(1,-\frac{1}{d};1-\frac{1}{d}; \frac{r^d_0}{r_h^d}\right)-1\right]=\frac{\el^2}{r_h}\frac{\pi}{d}\cot\frac{\pi}{d}-\frac{t_R}2\,.
\end{equation}
For the case  $d\geq2$, this equation can be solved only numerically.

Now let us first compute the bulk contribution from the Einstein-Hilbert action. By using the Einstein's equation, we can write this term as
\begin{equation}\label{intbulk1}
  I_{\text{bulk}}=\frac1{16\pi}\int\td^{d+1}x\sqrt{-g}\left[R+\frac{d(d-1)}{\el^2}\right]=-\frac{\Sigma_{d-1}d}{8\pi\el^2}\iint r^{d-1}\td r\td t\,.
\end{equation}
According to the right panel of Fig.~\ref{SAdS1}, the bulk term can be splited into four parts. In the region I,  for fixed $r$, the upper and inferior limits for $t$ in the integration \eqref{intbulk1} are given by the line equations $AB_1$ and $AE$, respectively. Thus we find
\begin{equation}\label{intbulk2a}
  I_{\text{bulk,I}}=-\frac{\Sigma_{d-1} d}{8\pi\el^2}\int_{r_h}^{r_m} r^{d-1}\td r\int^{r^*_m-r^*(r)}_{r^*(r)-r^*_m}\td t\,.
\end{equation}
At the region II, the coordinate $r$ can run from $r_0$ to $r_h$. For every given $r$, the upper and inferior limits for $t$ in the integration \eqref{intbulk1} are given by the line equations $AB$ and $BC$, respectively. Then we have,
\begin{equation}\label{intbulk2b2}
  I_{\text{bulk,II}}=-\frac{\Sigma_{d-1} d}{8\pi\el^2}\int_{r_0}^{r_h} r^{d-1}\td r\int^{r^*_m-r^*(r)}_{t_R+r^*(r)-r^*_m}\td t\,.
\end{equation}
For the region III and region IV, we have
\begin{equation}\label{intbulk2c}
  I_{\text{bulk,III}}=-\frac{\Sigma_{d-1} d}{8\pi\el^2}\int_{r_h}^{r_m} r^{d-1}\td r\int^{t_R+r^*_m-r^*(r)}_{t_R+r^*(r)-r^*_m}\td t=I_{\text{bulk,I}} \,,
\end{equation}
and
\begin{equation}\label{intbulk2d0}
  I_{\text{bulk,IV}}=-\frac{\Sigma_{d-1} d}{8\pi\el^2}\int_{0}^{r_h} r^{d-1}\td r\int^{t_R+r^*_m-r^*(r)}_{r^*(r)-r^*_m}\td t\,.
\end{equation}
%
%The parts I, III and IV can be computed by Eqs.~\eqref{intbulk2a} and \eqref{intbulk2c}.
Combining Eqs.~\eqref{intbulk2a}, \eqref{intbulk2b2}, \eqref{intbulk2c} and \eqref{intbulk2d0} we  find that
\begin{equation}\label{intbulk2d}
\begin{split}
  I_{\text{bulk}}(t_R)&=I_{\text{bulk,I}}+I_{\text{bulk,II}}+I_{\text{bulk,III}}+I_{\text{bulk,IV}}\\
  &=I_{\text{bulk},0}+\frac{\Sigma_{d-1} d}{8\pi\el^2}\int_0^{r_0}r^{d-1}\td r\int_{t_R+r^*(r)-r^*_m}^{r^*_m-r^*(r)}\td t\\
  %&=I_{\text{bulk},0}+\frac{\Sigma_{d-1} d}{4\pi\el^2}\int_0^{r_0}r^{d-1}\left[r^*_m-r^*(r)-\frac{t_R}2\right]\td r\\
  &=I_{\text{bulk},0}+\frac{\Sigma_{d-1}}{4\pi \el^2}\left[\left.r^d\left(r^*_m-r^*-\frac{t_R}2\right)\right|_0^{r_0}+\int_0^{r_0}r^{d-2}f^{-1}(r)\td r\right]\\
  &=I_{\text{bulk},0}+\frac{\Sigma_{d-1}}{4\pi\el^2}\int_0^{r_0}r^{d-2}f^{-1}(r)\td r\,.
  \end{split}
\end{equation}
Here $I_{\text{bulk},0}$ is the bulk on-shell action in the case that $t_R=0$.

Let us turn to the boundary terms. There are four null boundaries and a space-like boundary. As the normal vector fields are affinely parameterized, the only possible boundary terms for the null boundaries are the integration of $L_0$. Based on the results in the Ref.~\cite{Kim:2017lrw}, we see that
\begin{equation}\label{IlambdaBZT1}
\begin{split}
  I_{\mathcal{N}}(t_R)&= I_{\mathcal{N},0}+\frac{(d-1)\Sigma_{d-1}}{4\pi}\int^0_{r_0}r^{d-2}\left\{1+\ln\left[\frac{(d-1)\el}{r}\right]\right\}\td r\\
  &=I_{\mathcal{N},0}-\frac{\Sigma_{d-1}}{4\pi}\left\{\ln\left[\frac{(d-1)\el}{r_0}\right]+\frac{1}{(d-1)}\right\}r^{d-1}_0\,.%=I_{\lambda,0}-\frac{r_0}2-\frac{r_0}4\ln\left(\frac{\el^2}{r_0^2}\right)\,.
  \end{split}
\end{equation}
Here $I_{\mathcal{N},0}$ is the value of $I_{\mathcal{N}}(t_R)$ at $t_R=0$. The other boundary term comes from the space-like boundary $DE$. This boundary term is given by the GHY boundary term which reads
\begin{equation}\label{BTZGHY}
  I_{\text{GHY}}(t_R)=\frac{\Sigma_{d-1}r_h^{d}d}{16\pi\el^2}[t_R+2r^*_m-2r^*(0)]\,,
\end{equation}
based on Ref.~\cite{Lehner:2016vdi}.

Comparing with the case of $t_R=0$, we only have one new additional null-null joint term at $r_0$ which is
\begin{equation}\label{BTZjoint}
  I_{\text{joints}}(t_R)=I_{\text{joints},0}-\frac{r_0^{d-1}\Sigma_{d-1}}{8\pi}\ln[-r_0^2f(r_0)]\,,
\end{equation}
where $I_{\text{joints},0}$ is the value of $I_{\text{joints}}(t_R)$ at $t_R=0$.

Because of the time translation symmetry, the surface counterterms are the same as the cases of $t_R=0$. Thus the renormalized holographic complexity potential is
\begin{equation}\label{totalCAdS}
\begin{split}
  \mathcal{C}_{\text{A,ren}}(T,t_R)&=\mathcal{C}_{\text{A,ren},0}+\frac{\Sigma_{d-1}}{4\pi^2\hbar}\left\{\el^{-2}\int_0^{r_0}r^{d-2}f^{-1}(r)\td r+\frac{r_h^{d}(t_R-\Delta t_c)d}{4\el^2}\right.\\
  &\left.-\frac{r_0^{d-1}}2\left[\ln(-r_0^2f(r_0))+2\ln\left(\frac{(d-1)\el}{r_0}\right)+\frac{2}{d-1}\right]\right\}\,,
  \end{split}
\end{equation}
where $r_0$ is the function of $t_R$ and defined by the Eq.~\eqref{eqforr0} with $r_m\rightarrow\infty$. The value of $\mathcal{C}_{\text{A,ren},0}$ is given by Eq.~\eqref{CAregs0}.

%Now our mention is to find the minimal length of the curve to connect two points $(\epsilon,0)$ and $(x_0,s_0)$.

\subsubsection{Time dependnent complexity}
For convenience we choose $y=r_0/r_h$ and $x=(\el/r_h)^{d-1}$ as two free parameters of complexity potential. In this new coordinate,  the state $|\text{TFD}(T,t_R)\rangle$ is $|\text{TFD}(x,y)\rangle$.
%The corresponding zero temperature vacuum state then is given by $|\text{TFD}(0,0)\rangle$ for $d\geq3$.
Let us define a dimensionless renormalized complexity potential $G(x,y)$ such that,
\begin{equation}\label{defineGxs}
  \mathcal{C}_{\text{A,ren}}(x,y)=\frac{\el^{d-1}\Sigma_{d-1}}{4\pi^2\hbar}G(x,y)\,.
\end{equation}
%
%Take $a_d=\pi\frac{d-2}{d}\cot\frac{\pi}{d}$, we can easy to see that,
%Here $\text{Vol}_{d-1}:=\Sigma_{d-1}\el^{d-1}$ is the real volume of the dual boundary.
Comparing with Eq.~\eqref{totalCAdS} we find
\begin{equation}\label{totalCAdS2}
%\begin{split}
 G(x,y)=x^{-1}\left[\pi\frac{d-2}{d}\cot\frac{\pi}{d}+h(y)\right]\,,
 %\end{split}
\end{equation}
with
\begin{equation}\label{definhs}
  h(y)=\int_0^{y}s^{d-2}f_1^{-1}(s)\td s+\frac{f_2(y)d}{2}-\frac{y^{d-1}}2\left[\ln(-f_1(y))+2\ln(d-1)+\frac{2}{d-1}\right] \,,
\end{equation}
and
\begin{equation}\label{definf1f2}
  f_1(y)=1-\frac1{y^d},~~f_2(y)=\frac{\pi}{d}\cot\frac{\pi}{d}-\frac{1}{y}\left[_2F_1(1,-1/d;1-1/d;y^d)-1\right]\,.
\end{equation}
When $d\geq3$, the vacuum state is given by the parameter $x=0,y=0$. Suppose that $x=x_0$ and $y=y_0$ stand for an arbitrary TFD state. According to Eq.~\eqref{newcvca3a}, the minimal length connecting $(0,0)$ and $(x_0,y_0)$ is given by $|\mathcal{C}_{\text{A,ren}}(0,0)-\mathcal{C}_{\text{A,ren}}(x_0,y_0)|=|\mathcal{C}_{\text{A,ren}}(x_0,y_0)|$. However, when $d=2$, as Ref.~\cite{Chapman:2016hwi} suggested, the corresponding vacuum state is not the one of $r_h=0$. Instead, the vacuum state is given by $f(r)=1/\el^2+1/r^2$. The renormalized holographic complexity potential for this vacuum state is
\begin{equation}\label{CAregs0b}
  \frac{\pi\hbar\mathcal{C}_{\text{A,BTZ,vac}}}{\Sigma_{1}}=-\frac{\pi\el}2\,.
\end{equation}

Finally, we obtain the following complexity between the TFD state and its vacuum state
\begin{equation}\label{AdSCompCA1}
\begin{split}
  \mathcal{C}(|\text{TFD}(T,t_R)\rangle,|0\rangle)&=\frac{r_h^{d-1}\Sigma_{d-1}}{4\pi^2\hbar}[G(x_0,y_0)+2\pi^2\delta_{2,d}]\\
  &=\frac{d}{\pi^2(d-1)}[G(x_0,y_0)+2\pi^2\delta_{2,d}]\frac{M}{\hbar T}\,,
  \end{split}
\end{equation}
with the relationships $T=x_0^{1/(d-1)}d/(4\pi\el)$ and $t_R=2\el x^{-1/(d-1)}f_2(y_0)$. The absolute symbol has been dropped because the right side of Eq.~\eqref{AdSCompCA1} is always positive when $d>2$ (we confirmed it numerically from $d=2$ to $d=10$.). The growth rate can be obtained directly from this expression, which reads\footnote{When this paper was finished, Ref.~\cite{Carmi:2017jqz} appeared which also studied the complexity growth rate.  Our result Eq.~\eqref{dCdtAdS1} is the same as Eq.~(E.9) in Ref.~\cite{Carmi:2017jqz}.  }
\begin{equation}\label{dCdtAdS1}
\begin{split}
 &\frac{\td}{\td t_R}\mathcal{C}(|\text{TFD}(T,t_R)\rangle,|0\rangle)\\
 =&\frac{2M}{\pi\hbar}\left\{1+\frac{y_0^df_1(y_0)}{2\pi\hbar}\left[\ln(-f_1(y_0))+2\ln(d-1)\right)\right\}\,.
 \end{split}
\end{equation}
%

%This shows that the growth rate isn't monotonic with respective to $t_R$ and there is a maximal value which is larger than $2M/(\pi\hbar)$.

%
\begin{figure}
  \centering
  \includegraphics[width=.49\textwidth]{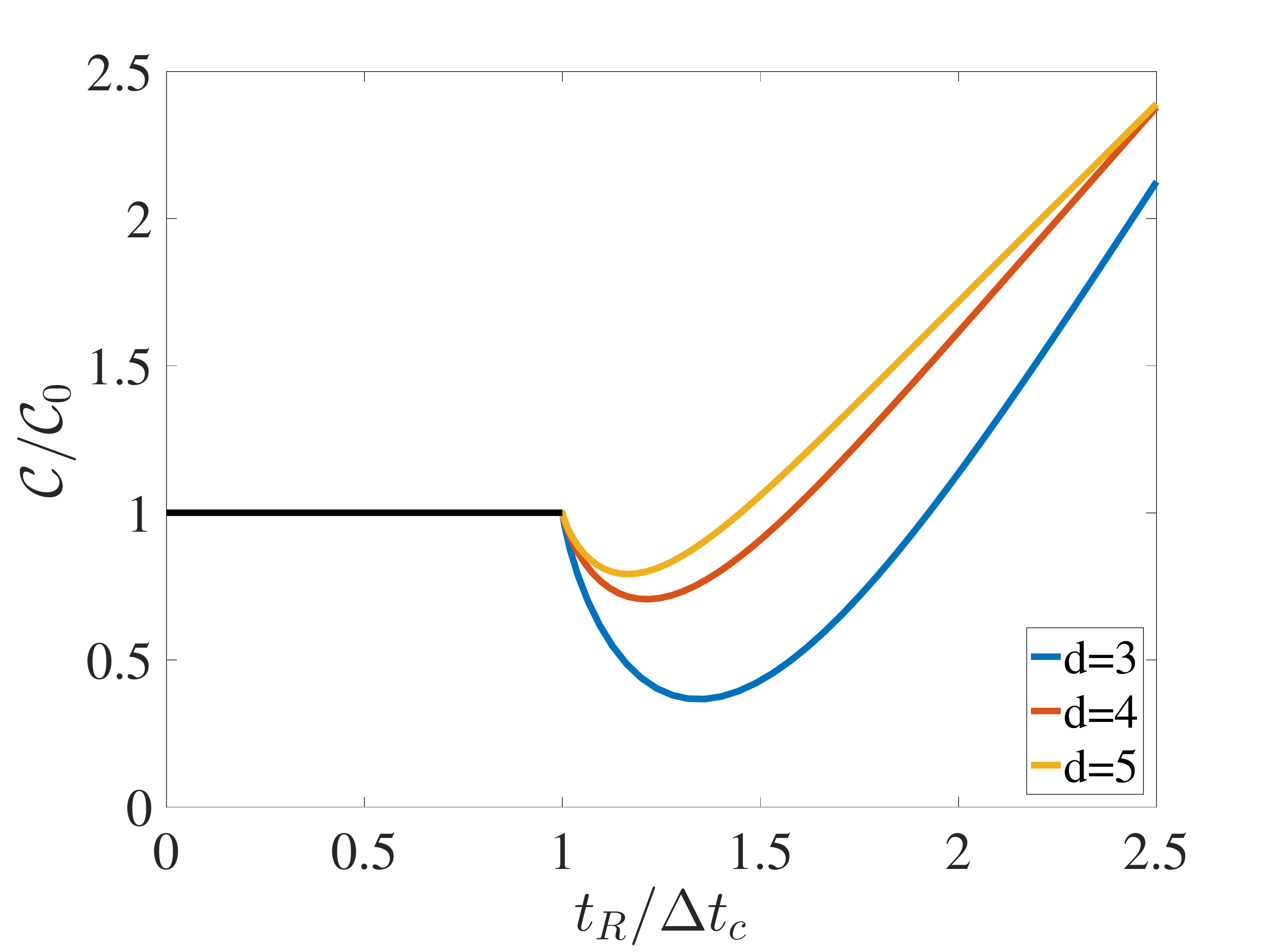}
   \includegraphics[width=.49\textwidth]{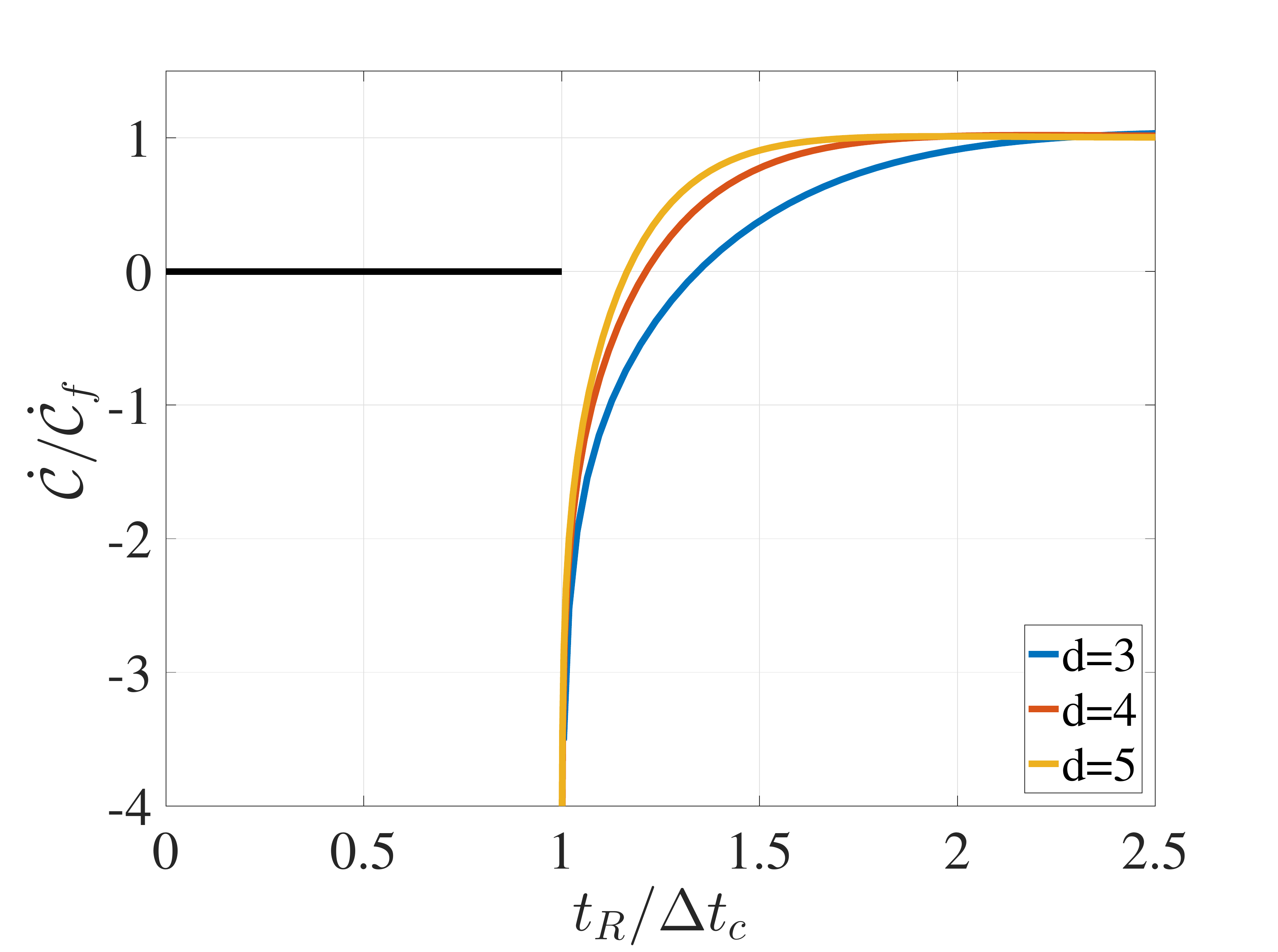}
  \caption{The complexity $\mathcal{C}(|\text{TFD}(0,t_R)\rangle,|0\rangle)$  and its growth rate  when $d>2$. $\mathcal{C}_0$ is the complexity when $t_R=\Delta t_c$ and $\dot{\mathcal{C}}_m=2M/\pi\hbar$. Higher dimensional cases give the similar results. }\label{FigAdS1}
\end{figure}
\begin{figure}
  \centering
  \includegraphics[width=.49\textwidth]{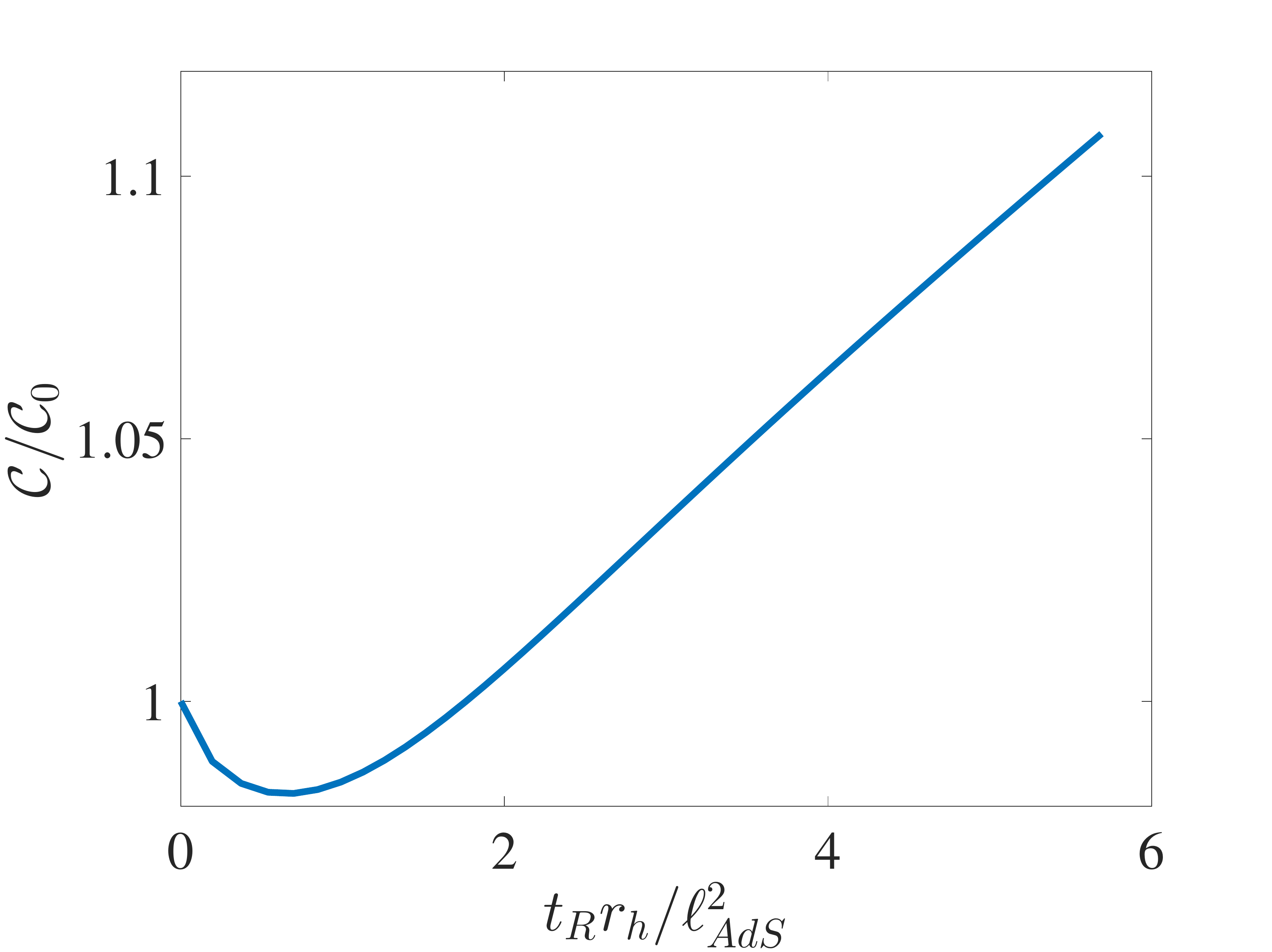}
   \includegraphics[width=.49\textwidth]{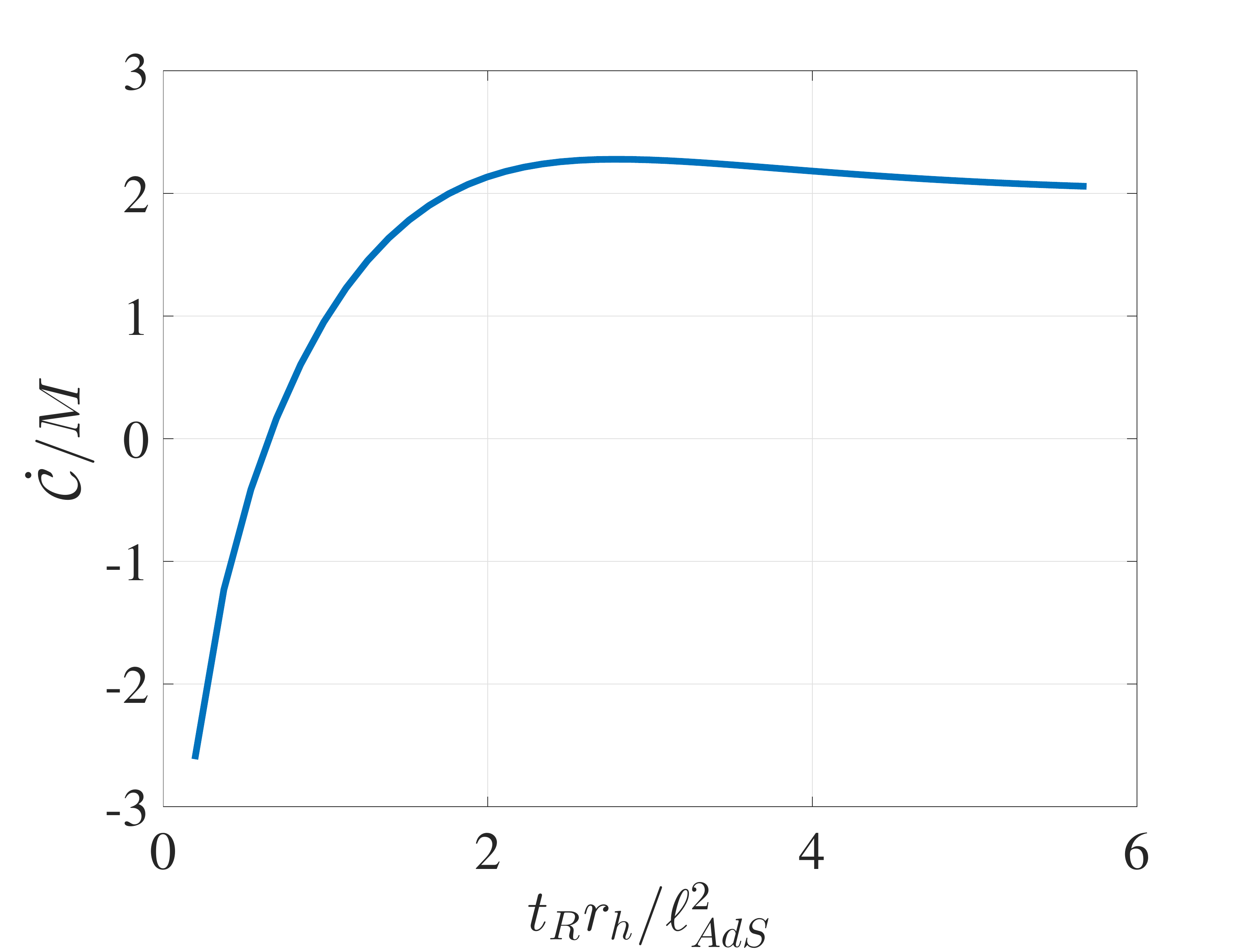}
  \caption{The complexity $\mathcal{C}(|\text{TFD}(0,t_R)\rangle,|0\rangle)$  and their growth rates for the BTZ black hole.  $\mathcal{C}_0$ is the complexity when $t_R=0$ and $\dot{\mathcal{C}}_m=2M/\pi\hbar$.}\label{FigBTZ1}
\end{figure}

The time evolution of the complexity $\mathcal{C}(|\text{TFD}(T,t_R)\rangle,|0\rangle)$ and its growth rate are shown in  Fig.~\ref{FigAdS1} and Fig.~\ref{FigBTZ1}. We find that the relationship between the complexity and $t_R$ is not  monotonic. When $t_R$ runs from $\Delta t_c$ to infinite, the value of complexity will first decrease and then increase, so there is a minimal value. For the case that $t_R\rightarrow\Delta t_c$, we have $y_0\rightarrow0^+$. Thus  Eq.~\eqref{dCdtAdS1} shows that
\begin{equation}\label{dCdtAdS2}
  \frac{\td}{\td t_R}\mathcal{C}(|\text{TFD}(T,t_R)\rangle,|0\rangle)\rightarrow-\infty,~~~\text{as}~t_R\rightarrow\Delta t_c\,.
\end{equation}
In the late limit $t_R\rightarrow\infty$, it saturates to the Lloyd's bound,
\begin{equation}\label{boundAdSa2}
  \lim_{t_R\rightarrow\infty}\frac{\td}{\td t_R}\mathcal{C}(|\text{TFD}(0,t_R)\rangle,|0\rangle)=\lim_{t_R\rightarrow\infty}\dot{\mathcal{C}}_{\text{A,ren}}=\lim_{y_0\rightarrow 1}\dot{\mathcal{C}}_{\text{A,ren}}=\frac{2M}{\pi\hbar}\,.
\end{equation}
From Fig. \ref{FigBTZ1} it is clear that the Lloyd's bound is violated in the intermediate and large time for the BTZ black hole ($d=2$), but it is not so clear if this is the case also for $d>2$ from Fig. \ref{FigAdS1}. To check it we consider the
the subleading term from Eq.~\eqref{dCdtAdS1} in the late time limit:
\begin{equation}\label{dCdtAdSs2}
 \frac{\td}{\td t_R}\mathcal{C}(|\text{TFD}(T,t_R)\rangle,|0\rangle)-\frac{2M}{\pi\hbar}=\frac{2M}{\pi\hbar}\frac{y_0^df_1(y_0)}{2\pi\hbar}\ln[-f_1(y_0)]+\mathcal{O}(y_0-1)\,.
\end{equation}
As $y_0\in(0,1)$, $f_1(y_0)=1-1/y_0^d<0$. In the late time limit, $y_0\rightarrow1^-$ and the first term in the right-side of Eq.~\eqref{dCdtAdSs2} dominant, which means that the subleading term is positive.
Thus the CA conjecture violates the Lloyd's bound slightly in the large time.

\subsection{CV conjecture}
\begin{figure}
	\begin{center}
		\begin{tikzpicture}
		
		\node (I)    at (0,0)   {};
		
		\path % left
		(I) +(135:4)  coordinate  (Iltop)
		+(-135:4) coordinate                     (Ilbot)
		+(0:0)   coordinate                  (Ilmid)
		+(45:4) coordinate        (Irtop)
		+(-45:4)   coordinate                  (Irbot)
		+(0:0)   coordinate                  (Irmid)
		;
		
		\draw  (Iltop) --
		node[pos=0.5, below, sloped]    {$\rho=\infty$}
		(Ilbot) --
		(Ilmid) --
		(Irtop) --
		node[pos=0.5, above, sloped]    {$\rho=\infty$}
		(Irbot) --
		node[midway, below, sloped]    {}
		(Irmid) --
		node[midway, below, sloped]    {$\rho=0$}
		(Iltop) -- cycle;
		
		% Squiggly lines
		\draw[decorate,decoration={snake,segment length=0.15cm,amplitude=0.05cm}] (Iltop)
		to[out=-15,in=+195,looseness=1.2]
		node[midway, above, inner sep=2mm] {$\kappa=\pi/d$}
		(Irtop);
		
		\draw[decorate,decoration={snake,segment length=0.15cm,amplitude=0.05cm}] (Ilbot)
		to[out=15,in=-195,looseness=1.2]
		(Irbot);
		
		% ER bridge
		\draw [blue,line width=.5mm]
		(-2.828,2.1)
		to[out=0,in=150,looseness=1.3]
		node[pos=-0.15,black]    {$\tilde{t}_B$}
		(-1.828,1.8);
		\draw [blue,line width=.5mm]
		(2.828,2.1)
		to[out=180,in=30,looseness=1.3]
		node[pos=-0.15,black]    {$\tilde{t}_B$}
		(1.828,1.8);
		\draw [blue,line width=.5mm]
		(-1.928,1.86) coordinate
		to[out=-30,in=210,looseness=1.36]
		node[black,midway, below]    {$\kappa_0$}
		(1.928,1.86) coordinate;
		\fill (0,1.1) circle (1.8pt);
		
		\draw[dashed,red,line width=.5mm]
		($(Iltop)!.0!(Ilbot)$) coordinate
		to[out=-40,in=220,looseness=1.4]
		node[black,midway, above]    {$\kappa_m=\pi/2d$}
		($(Irtop)!.0!(Irbot)$) coordinate;
		\draw [dashed,red,line width=.5mm]
		(-2.828,0) -- (2.828,0);
		\fill (0,1.34) circle (1.8pt);
		
		\end{tikzpicture}
		\caption{Extremal surfaces in the AdS black hole. For given time slices at the left and right boundary, the volume of extremal surface connecting these two time slices (blue curve) gives the holographic complexity potential. The upper red dotted line is for $\tilde{t}_B=\infty$ and the middle red dotted line is for $\tilde{t}_B=0$.}\label{FigAdS}
	\end{center}
\end{figure}
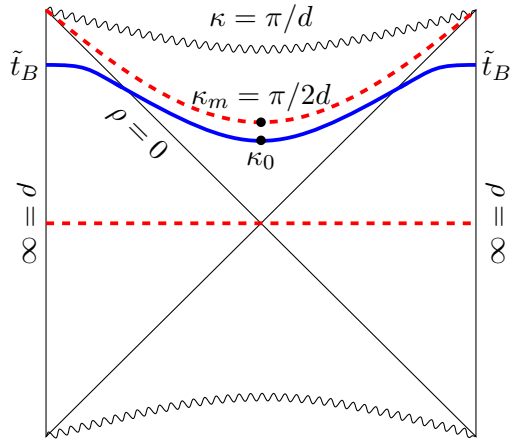
In this subsection, we compute the time-dependent complexity of the AdS$_{d+1}$ Schwarzschild planar black holes in the CV conjecture. Let us rewrite the metric~\eqref{metricBTZ} in the following form

\begin{equation}
	\td s^{2}=\el^2(-g^{2}(\rho)\td \tilde{t}^{2}+\td \rho^{2}+h^{2}(\rho)\sum_{i=1}^{d-1}\td \tilde{x}_i^2\label{eq:SAdSplanarBB}),
\end{equation}
where
\begin{equation}
	h(\rho)=\left(\cosh\frac{d\rho}{2}\right)^{2/d},\ \ g(\rho)=h(\rho)\tanh\frac{d\rho}{2}.
\end{equation}
Here we introduced dimensionless variables
$\tilde{t}=\frac{r_h}{\el^2}t$, $\tilde{x}_{i}=\frac{r_h}{\el^2}x_{i}$, $\tilde{r}=\frac{r}{r_h}$ and performed a coordinate transformation $\td \rho=\frac{\td \tilde{r}}{\tilde{r}\sqrt{1-1/\tilde{r}^{d}}}$.
%With out loss generality, we will also set $\ell=1$ in following.
The Penrose diagram is shown in Fig.~\ref{FigAdS}. Similarly to the CA case, the renormalized complexity potential only depends on $t_L+t_R$.

We can continue (\ref{eq:SAdSplanarBB}) into the interior region of Fig.~\ref{FigAdS} by setting $\rho=i\kappa$ and
$\tilde{t}_I=\tilde{t}+i\frac{\pi}{2}$. For the case $\tilde{t}_B\equiv \tilde{t}_R=\tilde{t}_L$, the maximal volume surface is given by the blue line in Fig.~\ref{FigAdS}. The upper red dotted line is for $\tilde{t}_B=\infty$ and the middle red dotted line is for $\tilde{t}_B=0$. The corresponding volume of this codimension-one surface is described by the following integration
\begin{align}\label{eq:vol}
	V= & \tilde{\Sigma}_{d-1}\el^d\int h(\rho)^{d-1}\sqrt{-g^{2}(\rho)+(\partial\rho/\partial \tilde{t})^{2}}\td \tilde{t},
\end{align}
where $\tilde{\Sigma}_{d-1}\equiv\int d^{d-1} \tilde{x}$ is the volume of the spatial geometry. The volume
can be maximized following \cite{Hartman:2013qma,MIyaji:2015mia,Sinamuli:2016rms}.

In principle, we should solve the Euler-Lagrangian equation of \eqref{eq:vol} to find $\rho(\tilde{t})$.
Alternatively, following \cite{Hartman:2013qma} we may find the first integral of the equation of motion of \eqref{eq:vol}. In other words, because the integrand of \eqref{eq:vol} is time independent the Hamiltonian is conserved:
\begin{align}\label{}
\mathcal{H}= & \frac{\partial \mathcal{L}}{\partial \rho'(\tilde{t})}-\mathcal{L}=const.\,,
\end{align}
which yields
\begin{align}\label{eq:energy}
\frac{g^2 h^{d-1}}{\sqrt{-g^2+(\partial\rho/\partial \tilde{t})^{2}}}= & i g_0 h_0^{d-1}\,,
\end{align}
where $h_0 := h(i\kappa_0)$ and $g_0:= g(i\kappa_0)$
with $\kappa_{0}$ ($0<\kappa_{0}<\frac{\pi}{2d}$) satisfying $\frac{\partial\kappa}{\partial \tilde{t}}| _{\kappa=\kappa_0}=0$.
From (\ref{eq:energy}), we can write the time $\tilde{t}_B$ in terms of $\kappa_0$
\begin{equation}
	\begin{split}
		\tilde{t}_B=&\int_{\epsilon}^{\kappa_{0}}\frac{\td\kappa}{\left(\cos\frac{d\kappa}{2}\right)^{\frac{2}{d}}\tan\frac{d\kappa}{2}\sqrt{1-\frac{\sin^{2}d\kappa}{\sin^{2}d\kappa_{0}}}}-\int_{\epsilon}^{\infty}\frac{\td\rho}{\left(\cosh\frac{d\rho}{2}\right)^{\frac{2}{d}}\tanh\frac{d\rho}{2}\sqrt{1+\frac{\sinh^{2}d\rho}{\sin^{2}d\kappa_{0}}}}\\
		=&\int_{0}^{\kappa_{0}}\left(
		\frac{\left(\cos\frac{d\kappa}{2}\right)^{-\frac{2}{d}}\cot\frac{d\kappa}{2}}{\sqrt{1-\csc^{2}d\kappa_{0}\sin^{2}d\kappa}}-\frac{\left(\cosh\frac{d\kappa}{2}\right)^{-\frac{2}{d}}\coth\frac{d\kappa}{2}}{\sqrt{1+\csc^{2}d\kappa_{0}\sinh^{2}d\kappa}}\right)\td\kappa\\
		&-\int_{\kappa_{0}}^{\infty}
		\frac{\left(\cosh\frac{d\rho}{2}\right)^{-\frac{2}{d}}\coth\frac{d\rho}{2}}{\sqrt{1+\csc^{2}d\kappa_{0}\sinh^{2}d\rho}}\td\rho\,.
	\end{split}
\end{equation}
Substituting (\ref{eq:energy}) into (\ref{eq:vol}), the maximum volume can be expressed in terms of the parameter $\kappa_{0}$,
\begin{equation}
	V=2\tilde{\Sigma}_{d-1} \el^d \left(\int_{0}^{\kappa_{0}}\frac{\left(\cos\frac{d\kappa}{2}\right)^{\frac{2(d-1)}{d}}}{\sqrt{\frac{\sin^{2}d\kappa_{0}}{\sin^{2}d\kappa}-1}}\td\kappa
	+\int_{0}^{\rho_{\infty}}\frac{\left(\cosh\frac{d\rho}{2}\right)^{\frac{2(d-1)}{d}}}{\sqrt{1+\frac{\sin^{2}d\kappa_{0}}{\sinh^{2}d\rho}}}\td\rho\right).
\end{equation}
Here we have introduced the UV cut off $\rho_{\infty}$, IR cut off $\epsilon$ and the factor 2 comes from the symmetry of Fig.~\ref{FigAdS}.

\begin{figure}
	\centering
	\includegraphics[width=.45\textwidth]{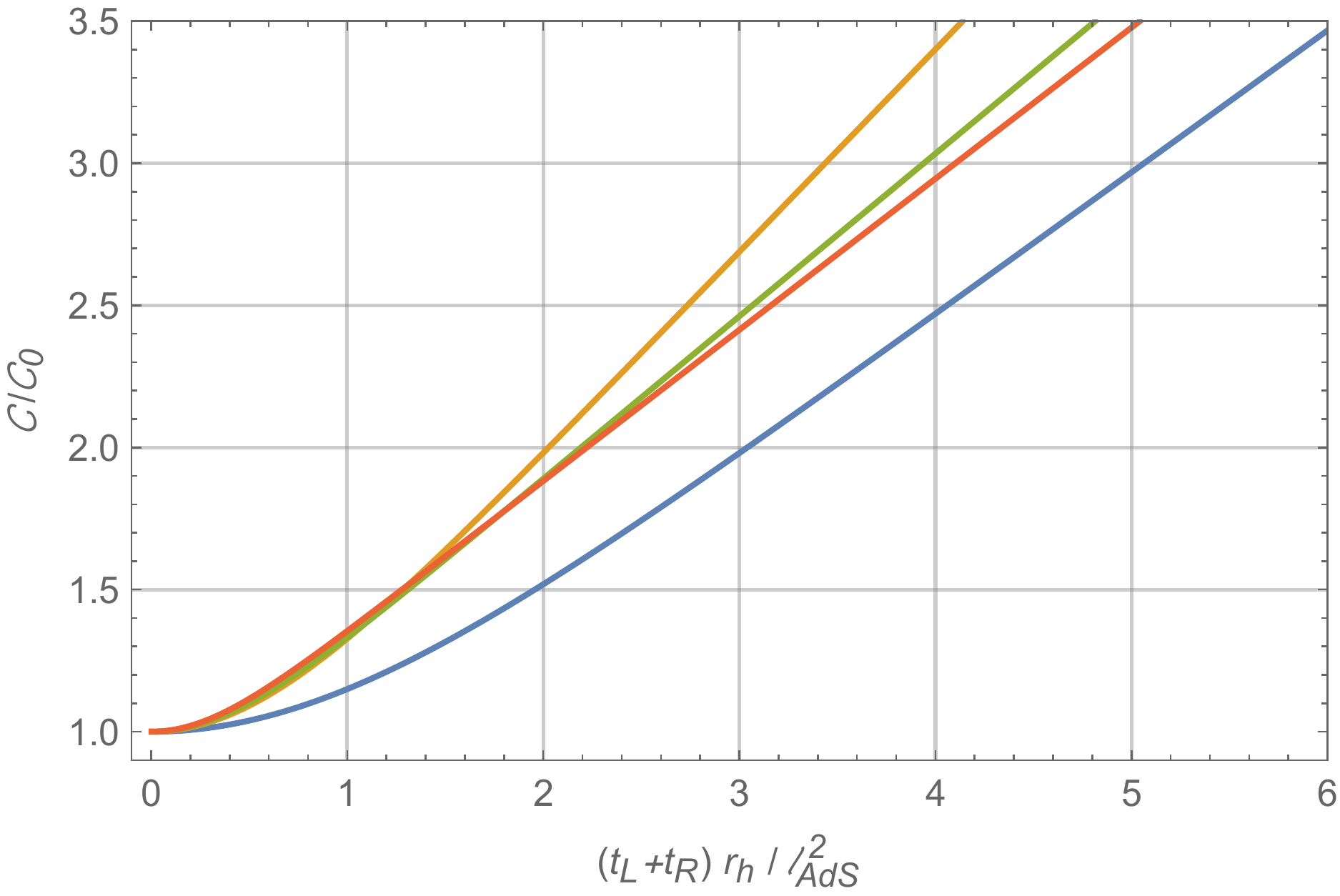}
	\includegraphics[width=.45\textwidth]{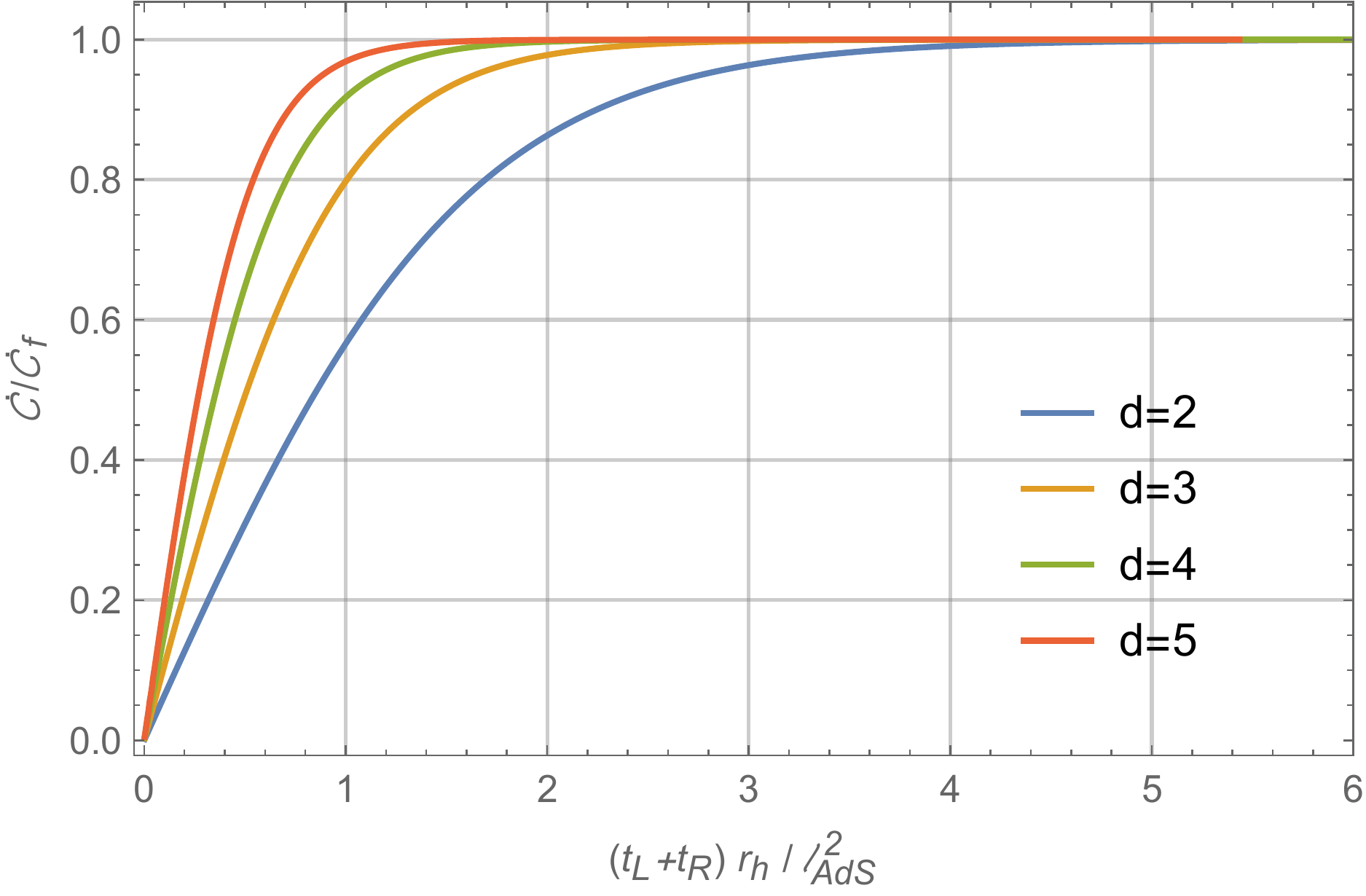}
	\caption{The values of $\mathcal{C}(|\text{TFD}(T,t_L+t_R)\rangle,|0\rangle)$ and its growth rate when $d=2,3,4,5$. The higher dimensions give similar results. $\mathcal{C}_0$ is the complexity when $t_L+t_R=0$ and $\dot{\mathcal{C}}_f$ is the Lioyd's bound of growth rate, which is given by Eq.~\eqref{dCdtAdS2CV}.}\label{Figrcv}
\end{figure}

%As we have explained in the introduction part, the complexity between a TDF state and its corresponding zero temperature vacuum state is given by the absolute value of renormalized complexity.
To evaluate the renormalized holographic complexity potantial, we will subtract the surface counterterms which were obtained for $d\ge2$ in Ref. \cite{Kim:2017lrw}:
\begin{equation}\label{}
	\begin{split}
		V^{(1)}_{\text{ct}}&=\frac{\el}{d-1}\int_Bd^{d-1} \tilde{x}\sqrt{\sigma}=\frac{\tilde{\Sigma}_{d-1} \el^d}{d-1}\left(\cosh\frac{d\rho_{\infty}}{2}\right)^{2(d-1)/d}\,,\\
		V^{(n)}_{\text{ct}}&=0,~~~~n>1,
	\end{split}
\end{equation}
where $\sigma$ is the induced metric of the time slice on the boundary. Hence the renormalized holographic complexity potential can be written as
\begin{equation}\label{rcv}
	\begin{split}
		&\mathcal{C}_{\text{V,ren}}
		=\frac{1}{\ell}\lim_{\delta\rightarrow0}(V-2V^{(1)}_{\text{ct}})\\
		%&=
		%\int_{0}^{\kappa_{0}}\frac{\left(\cos\frac{d\kappa}{2}\right)^{\frac{2(d-1)}{d}}}{\sqrt{\frac{\sin^{2}d\kappa_{0}}{\sin^{2}d\kappa}-1}}d\kappa
		%+\int_{0}^{\infty}\left(\frac{\left(\cosh\frac{d\rho}{2}\right)^{\frac{2(d-1)}{d}}}{\sqrt{1+\frac{\sin^{2}d\kappa_{0}}{\sinh^{2}d\rho}}}-\left(\cosh\frac{d\rho}{2}\right)^{\dfrac{d-2}{d}}\sinh\frac{d\rho}{2}\right)d\rho-\frac{1}{d-1}\\
		&= \frac{2\tilde{\Sigma}_{d-1}\el^d}{\ell}\left( \int_{0}^{\kappa_{0}}\frac{\left(\cos\frac{d\kappa}{2}\right)^{\frac{2(d-1)}{d}}}{\sqrt{\frac{\sin^{2}d\kappa_{0}}{\sin^{2}d\kappa}-1}}\td\kappa
		+ \int_{0}^{\infty}\left(\frac{\left(\cosh\frac{d\rho}{2}\right)^{\frac{2(d-1)}{d}}}{\sqrt{1+\frac{\sin^{2}d\kappa_{0}}{\sinh^{2}d\rho}}}-\frac{\cosh\frac{d\rho}{2}}{\left(\sinh\frac{d\rho}{2}\right)^{\dfrac{d}{d-2}}}\right)\td\rho
		\right)\,.
	\end{split}
\end{equation}
As in the CA conjecture, the renormalized holographic complexity potential of Schwarzschild AdS black holes at the zero temperature limit are all zeros, the complexity between $|\text{TFD}(T,t_L+t_R)\rangle$ and $|\text{TFD}(0,0)\rangle$ then is\footnote{We have substituted $t_L + t_R$ for $2 t_B$ for comparison with the results obtained by other methods.}
\begin{equation}\label{CVcomple1}
	\mathcal{C}(|\text{TFD}(T,t_L+t_R)\rangle,|0\rangle)=\mathcal{C}_{\text{V,ren}}\,.
\end{equation}
However, for the BTZ black hole, the vacuum state is not the one of zero horizon. We have to choose the solution $f(r)=1/\el^2+1/r^2$ for the vacuum state. Then the renormalized holographic complexity potential of this vacuum state is given by $\mathcal{C}_{V,\text{BTZ,vac}}=-4\pi\el^2/\ell$. Thus we have the following complexity for the BTZ black hole
\begin{equation}\label{CVcomple2}
	\mathcal{C}(|\text{TFD}(T,t_L+t_R)\rangle,|0\rangle)=\mathcal{C}_{\text{V,ren}}+4\pi\el^2/\ell\,.
\end{equation}
%
%
%%
%\begin{equation}\label{rcv2}
%  \mathcal{C}(|\text{TFD}(T,t_R)\rangle,|0\rangle)=\ell^{-1}\mathcal{C}_{\text{V,ren}}\,.
%\end{equation}
%%
%
Combining Eqs.~\eqref{CVcomple1} and \eqref{CVcomple2} we have an expression
\begin{equation}\label{CVcomple0}
	\mathcal{C}(|\text{TFD}(T,t_L+t_R)\rangle,|0\rangle)=\mathcal{C}_{\text{V,ren}}+4\pi\delta_{2,d}\el^2/\ell\,.
\end{equation}
The time evolution of the complexity $\mathcal{C}(|\text{TFD}(T,t_L+t_R)\rangle,|0\rangle)$ and its growth rate are shown in  Fig.~\ref{Figrcv} where the relationship between the complexity and $t_L+t_R$ is monotonic contrary to the CA case.

In the early time limit ($\tilde{t}_B\rightarrow0$ or $\kappa_{0}\rightarrow0$
), we have
%{\color{red}
\begin{equation}
	\tilde{t}_B=-\frac{\sin d\kappa_{0}}{2}\int_{0}^{\rho_{\infty}}\frac{\td\rho}{\left(\cosh\frac{d\rho}{2}\right)^{\frac{2}{d}}\sinh^{2}\frac{d\rho}{2}},
\end{equation}
so the complexity $\mathcal{C}(|\text{TFD}(T,t_L+t_R)\rangle,|0\rangle)$ can be written as
\begin{align}
	&\mathcal{C}(|\text{TFD}(T,t_L+t_R)\rangle,|0\rangle)=\mathcal{C}_{\text{V,ren}}+4\pi\delta_{2,d}\el^2\nonumber\\
	&\quad=\frac{\tilde{\Sigma}_{d-1}\el^d}{\ell}\left(
	\frac{\sqrt{\pi}(d-2)\Gamma(1+\frac{1}{d})}{(d-1)\Gamma(\frac{1}{2}+\frac{1}{d})}
	+\frac{r_h^2d^2\Gamma(\frac{1}{2}+\frac{1}{d})}{8\el^4\sqrt{\pi}\Gamma(\frac{1}{d})}(t_L+t_R)^2+\dots
	\right)+4\pi\delta_{2,d}\el^2/\ell.
\end{align}
%
%({\color{blue}what is ``$\ell$''?})
At  $t_L+t_R=0$,
\begin{equation}\label{CVComMass}
	\mathcal{C}(|\text{TFD}(T,0)\rangle,|0\rangle)= \frac{d\sqrt{\pi}(d-2)\Gamma(1+\frac{1}{d})}{\pi^2(d-1)\Gamma(\frac{1}{2}+\frac{1}{d})}\frac{M}{\hbar T}+\frac{(d-1)\el}{\pi\hbar}\delta_{2,d}\,,
\end{equation}
where we use Eq.~\eqref{massAdS} and take the length scale $\ell/\el=4\pi^2\hbar/(d-1)$.
%The particular case of $d=2$ is special, which is described by the non-rotational BTZ black hole.
%

In the late time limit ($\tilde{t}_B\rightarrow\infty$ or $\kappa_{0}\rightarrow\kappa_m=\frac{\pi}{2d}$
), the renormalized complexity \eqref{rcv} becomes
\begin{equation}
	\begin{split}
		&\mathcal{C}(|\text{TFD}(T,t_L+t_R)\rangle,|0\rangle) =\mathcal{C}_{\text{V,ren}} \\
		&\quad =\frac{\tilde{\Sigma}_{d-1}\el^d}{\ell}\left(
		\frac{r_h}{2\el^2}(t_L+t_R) -\int_{\epsilon}^{\frac{\pi}{2d}}\frac{\cos d\kappa~\td\kappa} {\left(\cos\frac{d\kappa}{2}\right)^{2/d}\tan\frac{d\kappa}{2}} +\int_{\epsilon}^{\rho_{\infty}}\frac{\coth\frac{d\rho}{2}~\td\rho}{\left(\cosh\frac{d\rho}{2}\right)^{2/d}}
		-\dfrac{2}{d-1}\right)\\
		&\quad =\frac{\tilde{\Sigma}_{d-1}\el^d}{\ell}\left(\frac{r_h}{2\el^2}(t_L+t_R)+\text{finite term}\right).
	\end{split}
\end{equation}
Thus the complexity growth rate in the late time limit is given by
\begin{equation}\label{dCdtAdS2CV}
	\lim_{t_L+t_R\rightarrow\infty}\frac{\td}{\td (t_L+t_R)}\mathcal{C}(|\text{TFD}(T,t_L+t_R)\rangle,|0\rangle)=\frac{8\pi\el M}{\ell(d-1)}\,,
\end{equation}
where we use Eq.~\eqref{massAdS}.
If we take the length scale $\ell/\el=4\pi^2\hbar/(d-1)$ we find that the Lloyd's bound is satisfied in the CV conjecture:
\begin{equation}\label{dCdtAdS2CV}
	\frac{\td}{\td (t_L+t_R)}\mathcal{C}(|\text{TFD}(T,t_L+t_R)\rangle,|0\rangle)<\frac{2M}{\pi\hbar}\,.
\end{equation}
Numerical results show that this is the maximum value of the growth rate at all time, which is different from the CA conjecture.

Let us make a comparison for the complexity growth rates between the CA and CV conjectures. From Figs.~\ref{FigAdS1}, \ref{FigBTZ1} and \ref{Figrcv} we see that at early time, two conjectures give different results.
In the CA conjecture, we see that the complexity between the TFD state and the vacuum state does not change until $t_R+t_L=\Delta t_c$ for $d>2$. When $t>\Delta t_c$, the CA conjecture predicts the complexity will decrease first and then increase.  The growth rate at $t=\Delta t_c$ is negative infinite. In the CV conjecture, we see that the complexity between the TFD state and the vacuum state always increase with the order of $t^2$ when $t$ is small.  In  the late time limit, two conjectures predict the complexity will increase linearly in $t$ and the slope is proportional to total mass $M$. However, in the large time region, the CA conjecture will approach to $2M$ from a larger value, so it violates the Lloyd's bound. If we choose the length scale $\ell=4\pi^2\hbar\el/(d-1)$, the CV conjecture satisfies the Lloyd's bound at all time  and saturates to the Lloyd's bound in the late time limit.
%However, there is still a difference at late time limit. For CA conjecture, will see that $\dot{\mathcal{C}}_{\text{A,ren}}(t)$ will decrease and leads to the constant. For CV conjecture, we can see that $\dot{\mathcal{C}}_{\text{V,ren}}(t)$ will increase and leads to the constant.

\section{Time dependent complexity of the TFD states:  field theory approach}\label{fromQTF}

In this section we compute the complexity by the field theoretic methods proposed by  Refs.~\cite{Chapman:2017rqy,Yang:2017nfn}. One is the FS method~\cite{Chapman:2017rqy} based on the Fubini-Study metric and the other is the FG method~\cite{Yang:2017nfn} based on the Finsler geometry. As a common basis of two methods we start with constructing a time-dependent TFD state for free field theory explicitly.

\subsection{Time evolution of the TFD states}\label{tTFD}
Both in the FS and FG methods, a crucial step is to find the transformation from vacuum state to a TFD state. We will follow the method proposed by Ref.~\cite{Yang:2017nfn}. Let us consider a bosonic Hilbert space $\mathcal{H}$ and the occupation number representation. Suppose that $\hat{a}_{\vec{k}_i}$ and $\hat{a}^\dagger_{\vec{k}_i}$ are the annihilation and creation operators, which can annihilate or create a particle of momentum $\vec{k}_i$. The particle number density operator at momentum $\vec{k}_i$ is defined as
\begin{equation}\label{Nakak1}
\hat{N}_{\vec{k}_i}:=\hat{a}^\dagger_{\vec{k}_i}\hat{a}_{\vec{k}_i}\,.
\end{equation}
As the particle number density operators for different momentum commute each other, their common eigenstates form a complete basis in the Hilbert space $\mathcal{H}$. Let us assume the momentum is discrete and introduce the notation,
\begin{equation}\label{Nbasis}
  \prod_{i}|n_i,\vec{k}_i\rangle:=|n_0,\vec{k}_0\rangle|n_1,\vec{k}_1\rangle|n_2,\vec{k}_2\rangle\cdots
\end{equation}
to stand for one common eigenstate for all the particle number operators. Here the product includes all the possible momentum values. Then any state $|\psi\rangle\in\mathcal{H}$ can be written in the following form
\begin{equation}\label{cn0n1n2}
  |\psi\rangle=\sum_{n_0,n_1,\cdots=0}^\infty c_{n_0n_1\cdots}|n_0,\vec{k}_0\rangle|n_1,\vec{k}_1\rangle\cdots\,.
\end{equation}

To construct a TFD state, one method is using a bogoliubov transformation from the vacuum state defined by any chosen annihilation operators~\cite{Yang:2017nfn}. Let us first decompose the Hilbert space $\mathcal{H}=\mathcal{H}_L\times\mathcal{H}_R$ and define two groups of annihilation and creation operators \{$\hat{a}_{\vec{k}_i}^L, \hat{a}_{\vec{k}_i}^R$\} and  \{$\hat{a}_{\vec{k}_i}^{L\dagger}, \hat{a}_{\vec{k}_i}^{R\dagger}$\}, which define a vacuum state $|0\rangle$:
\begin{equation}\label{defineA}
  \hat{a}_{\vec{k}_i}^L|0\rangle=\hat{a}_{\vec{k}_i}^R|0\rangle=0,~~~\forall~\vec{k}_i\,.
\end{equation}

However, the decomposition $\mathcal{H}=\mathcal{H}_L\times\mathcal{H}_R$ is not unique. One can choose  another decomposition such that $\mathcal{H}=\mathcal{H}_D\times\mathcal{H}_U$ with the annihilation and creation operators \{$\hat{b}_{\vec{k}_i}^D, \hat{b}_{\vec{k}_i}^U$\} and  \{$\hat{b}_{\vec{k}_i}^{D\dagger}, \hat{b}_{\vec{k}_i}^{U\dagger}$\}. The annihilation operators \{$\hat{b}_{\vec{k}_i}^D, \hat{b}_{\vec{k}_i}^U$\} also define a vacuum state $|B\rangle$ such that $\hat{b}_{\vec{k}_i}^U|B\rangle=\hat{b}_{\vec{k}_i}^D|B\rangle=0$ for $\forall~\vec{k}_i$. In general, two kinds of decompositions can have no special relationship. However, if we demand that they satisfy the following relationship
\begin{equation}\label{relab}
\left[\begin{matrix}
\hat{b}^D_{\vec{k}}\\
\hat{b}^{U\dagger}_{\vec{k}}
\end{matrix}\right]=c_{\vec{k}}\left[\begin{matrix}
1&-e^{-\pi\omega_{\vec{k}}/a}\\
-e^{-\pi\omega_{\vec{k}}/a}&1\end{matrix}\right]
\left[\begin{matrix}
\hat{a}^L_{\vec{k}}\\
\hat{a}^{R\dagger}_{\vec{k}}
\end{matrix}\right] \,,
\end{equation}
for a normalization factor $c_{\vec{k}}$ which makes the operators \{$\hat{b}_{\vec{k}_i}^D, \hat{b}_{\vec{k}_i}^U$\} and  \{$\hat{b}_{\vec{k}_i}^{D\dagger}, \hat{b}_{\vec{k}_i}^{U\dagger}$\} to satisfy the canonical commutation relation, Ref.~\cite{Yang:2017nfn} and  Ref.~\cite{Crispino:2007eb} have  proven that the vacuum state $|B\rangle$ can be expressed by
\begin{equation}\label{relBnk}
  |B\rangle\propto\prod_{\vec{k}_i}\sum_{n=0}^\infty e^{-\pi n\omega_{\vec{k}_i}/a}|n,\vec{k}_i\rangle_L|n,\vec{k}_i\rangle_R\,.
\end{equation}
There is a non-unitary operator
\begin{equation}\label{relaUaa}
  \hat{U}^\dagger_a:=\prod_{\vec{k}_i}\exp\left(e^{-\pi\omega_{\vec{k}_i}/a}\hat{a}^{R\dagger}_{\vec{k}_i}\hat{a}^{L\dagger}_{\vec{k}_i}\right) =\exp\left[\int\td^{d-1}ke^{-\pi\omega_{\vec{k}}/a}\hat{a}^{R\dagger}(\vec{k})\hat{a}^{L\dagger}(\vec{k})\right] \,,
\end{equation}
which can convert the vacuum state $|0\rangle$ into the $|B\rangle$, i.e., $|B\rangle\propto\hat{U}^\dagger_a|0\rangle$. In the second equality, the discrete form has been converted into a continuous form.\footnote{Ref.~\cite{Yang:2017nfn} proves that $\hat{U}^\dagger_a$ has a unitary partner,
\begin{equation}\label{defineGa}
  \hat{G}_a:=\exp\left\{\int\text{arctanh}e^{-\pi\omega_{\vec{k}}/a}[\hat{a}^{R\dagger}(\vec{k})\hat{a}^{L\dagger}(\vec{k})-\hat{a}^{R}(\vec{k})\hat{a}^{L}(\vec{k})]\td^{d-1}k\right\}\,,
\end{equation}
which can also realize $|B \rangle\propto\hat{G}_a|0\rangle$.}

In order to prove the state $|B\rangle$ is a TFD state, the easiest way is to find the reduced density matrix in the projected Hilbert space $\mathcal{H}_L$ or $\mathcal{H}_R$. Ref.~\cite{Crispino:2007eb} has shown that the state in Eq.~\eqref{relBnk} has the following reduced density matrix
\begin{equation}\label{inducedrho}
  \hat{\rho}_L=\hat{\rho}_R=\frac1Z\prod_{\vec{k}_i}\sum_{n=0}^\infty\exp(-2\pi n\omega_{\vec{k}_i}/a)|n,\vec{k}_i\rangle\langle n,\vec{k}_i|\,,
\end{equation}
where the factor $1/Z$ insures that $\text{Tr}\hat{\rho}_L=\text{Tr}\hat{\rho}_R=1$. We see that this is the density matrix for the system of free bosons with temperature $T=a/2\pi$. Thus, the projected states of $|B\rangle$ in Hilbert space $\mathcal{H}_L$ and $\mathcal{H}_R$ are two thermofield state. This shows that $|B\rangle$ is a TFD state with temperature $T=a/2\pi$.

The time dependent TFD state is given by Eq.~\eqref{timesate1} so we have
\begin{equation}\label{timeTFD2}
  |\text{TFD}(t_L,t_R)\rangle=\exp[-i(\hat{H}_Lt_L+\hat{H}_Rt_R)]|B\rangle\propto\exp[-i(\hat{H}_Lt_L+\hat{H}_Rt_R)]\hat{U}_a^\dagger|0\rangle\,.
\end{equation}
The Hamiltonian $H_R$ and $H_L$ depend on the dynamic of dual boundary fields. For the free bosons, the Hamiltonian can be expressed by the creation and annihilation operators in the following way\footnote{The zero point energy has been neglected, as it only contribute a constant factor on the state. }
\begin{equation}\label{Hamilton1}
  \hat{H}_Lt_L+\hat{H}_Rt_R=\int\td^{d-1}k\omega_{\vec{k}}(\hat{N}_{\vec{k}}^Rt_R+\hat{N}_{\vec{k}}^Lt_L)\,.
\end{equation}
Although in general we cannot find a function $f(a,\vec{k})$ such that $\exp[-i(\hat{H}_Lt_L+\hat{H}_Rt_R)]\hat{U}_a^\dagger=\exp\left[\int\td^{d-1}kf(a,\vec{k})\hat{a}^{R\dagger}(\vec{k})\hat{a}^{L\dagger}(\vec{k})\right]$, we can find a function $f(a,\vec{k})$ satisfying
\begin{equation}\label{timeTFD3}
  |\text{TFD}(t_L,t_R)\rangle\propto\exp\left[\int\td^{d-1}kf(a,\vec{k})\hat{a}^{R\dagger}(\vec{k})\hat{a}^{L\dagger}(\vec{k})\right]|0\rangle\,.
\end{equation}

To see this, let us plug  Eqs.~\eqref{Hamilton1} and \eqref{relBnk} into Eq.~\eqref{timeTFD2} in the discrete form. Thus we have
\begin{equation}\label{timeTFD4}
\begin{split}
  |\text{TFD}(t_L,t_R)\rangle&\propto\prod_{\vec{k}_i}\sum_{n=0}^\infty e^{-\pi n\omega_{\vec{k}_i}/a}\exp[-i(\hat{N}_{\vec{k}}^Rt_R+\hat{N}_{\vec{k}}^Lt_L)]|n,\vec{k}_i\rangle_L|n,\vec{k}_i\rangle_R\\
  &\propto\prod_{\vec{k}_i}\sum_{n=0}^\infty e^{-n\omega_{\vec{k}_i}[\pi/a+i(t_R+t_L)]}|n,\vec{k}_i\rangle_L|n,\vec{k}_i\rangle_R \,.
  \end{split}
\end{equation}
Now converting it into the continuous form, we obtain that
\begin{equation}\label{timeTFD5}
  |\text{TFD}(t_L,t_R)\rangle\propto\hat{U}^\dagger_a(t_L,t_R)|0\rangle\,.
\end{equation}
with the time dependent non-unitary operator\footnote{Note that $\exp[-i(\hat{H}_Lt_L+\hat{H}_Rt_R)]\hat{U}_a^\dagger\neq\hat{U}^\dagger_a(t_L,t_R)$. The non-unitary operator $\hat{U}^\dagger_a(t_L,t_R)$ has a unitary partner:
\begin{equation}\label{defineGa2}
  \hat{G}_a(t_L,t_R):=\exp\left\{\int[r_{\vec{k}}\hat{a}^{R\dagger}(\vec{k})\hat{a}^{L\dagger}(\vec{k})-r_{\vec{k}}^*\hat{a}^{R}(\vec{k})\hat{a}^{L}(\vec{k})]\td^{d-1}k\right\}\,,
\end{equation}
Thus the time evolution of the TFD state can be generated by two ways: $ |\text{TFD}(t_L,t_R)\rangle\propto\hat{U}^\dagger_a(t_L,t_R)|0\rangle\propto \hat{G}_a(t_L,t_R)|0\rangle$.}
\begin{equation}\label{timeTFD5b}
  \hat{U}^\dagger_a(t_L,t_R):=\exp\left[\int\td^{d-1}ke^{-\omega_{\vec{k}}[\pi/a+i(t_R+t_L)]}\hat{a}^{R\dagger}(\vec{k})\hat{a}^{L\dagger}(\vec{k})\right]\,.
\end{equation}
This shows that the function in Eq.~\eqref{timeTFD3} is
\begin{equation}\label{funcfak}
  f(a,\vec{k})=e^{-\omega_{\vec{k}}[\pi/a+i(t_R+t_L)]}\,.
\end{equation}
We see that the time dependent TFD state only depends on $t_L+t_R$. For later use we also define
\begin{equation}\label{defrk}
  r_{\vec{k}}:= \text{arctanh}[f(a,\vec{k})]=\text{arctanh}e^{-\omega_{\vec{k}}[\pi/a+i(t_R+t_L)]}\,.
\end{equation}
Eqs. \eqref{funcfak} and \eqref{defrk} will play  crucial roles when we compute the complexity growth rate in the next subsections.

%As the complexity of state depends on the reference state, there are two different ways to define the growth rate.

%\subsection{complexity growth rate}
\subsection{Fubini-Study (FS) metric}\label{FSmetric0}
Let us first use the method proposed by Ref.~\cite{Chapman:2017rqy} to compute the complexity between $|\text{TFD}(t_L,t_R)\rangle$ and the zero temperature limit vacuum state $|0\rangle$. This method is based on the  Fubini-Study metric (see the appendix~\ref{FSmetric} for some basic introduction and  refer to Ref.~\cite{bengtsson2006geometry} for details) and unitary transformations. For a generator set $E=\{M^1,M^2,\cdots\}$, the tangent anti-Hermitian operator $\hat{T}$ can be decomposed as
\begin{equation}\label{decompT}
  \hat{T}(s)=\sum_IY_I(s)M^I\,.
\end{equation}
This tangent operator can generate a unitary operator by a time order exponential map
\begin{equation}\label{mapus1}
  \hat{O}(s):=\overleftarrow{P}\exp\left[\int_{0}^s\hat{T}(\tilde{s})\td \tilde{s}\right]\,,
\end{equation}
where $\overleftarrow{\mathcal{P}}$ denotes a time ordering such that the tangent operator at earlier times is applied to the state first. This $s$-dependent operator can induce a curve $c:[0,1]\mapsto\mathcal{H}$ such that
\begin{equation}\label{curvepsi1}
  c(s):=\hat{O}(s)|R\rangle\,, \qquad c(0)=|R\rangle \,,  \qquad c(1)=|T\rangle\,.
\end{equation}
This curve is determined by the generator set ($E$) and the coefficients ($Y_I$) of tangent operator which are shown in Eq.~\eqref{decompT}. Let us assume the image of the curve is $|\psi(s)\rangle$. We can compute the length of this curve by  the Fubini-Study metric
\begin{equation}\label{lenght1}
  \mathcal{L}[c]:=\int_0^1[||\partial_s|\psi(s)\rangle|^p-|\langle\psi(s)|\partial_s|\psi(s)\rangle|^p|]^{1/p}\td s\,.
\end{equation}
This paper will focus on the $L^1$ normal, i.e., $p=1$ because it was shown that $p=1$ case leads that the complexity density resembles the divergence structure of holographic complexity~\cite{Chapman:2017rqy}.

The complexity between the states $|T\rangle$ and $|R\rangle$ is given by the following optimization problem,
\begin{equation}\label{compFS}
\begin{split}
  \mathcal{C}(|T\rangle,|R\rangle):=&\min\left\{\mathcal{L}[c]~|\forall c:[0,1]\mapsto\mathcal{H}, \ c(0)=|R\rangle, \ c(1)=|T\rangle, \right.\\
  &\left.\text{and}~\exists\{Y^I\}~\text{such that}~\frac{\td}{\td s}c(s)=\sum_IY_I(s)M^Ic(s)\right\} \,.
  \end{split}
\end{equation}
By this definition, the choice of generator set $E$ may affect the complexity between two states. So far the generator set $E$ is arbitrary and there may be many possible choices. Finding the complexity in a very general generator set seems to be a too mathematical and technical problem.  However, in this subsection, we want to compute the complexity between $|\text{TFD}(t_L,t_R)\rangle$ and $|0\rangle$ which are related by the operators $ \hat{U}_a^\dagger(t_L,t_R)$.  Because the TFD states can be generated by some generators which form a su(1,1) Lie algebra, as will be shown in \eqref{sampleU1}, we choose,  as a minimal nontrivial generator set,
\begin{equation}\label{Eset1}
  E_L=\bigcup_{\vec{k}}\{\hat{L}^{(\vec{k})}_+,\hat{L}^{(\vec{k})}_-,\hat{L}_0^{(\vec{k})}\} \,,
\end{equation}
with
\begin{equation}\label{threeLs}
\begin{split}
&\hat{L}_+^{(\vec{k})}:=\hat{a}^{R\dagger}(\vec{k})\hat{a}^{L\dagger}(\vec{k})\,, \\
&\hat{L}_-^{(\vec{k})}:=\hat{a}^{R}(\vec{k}) \hat{a}^{L}(\vec{k})\,, \\
& \hat{L}_0^{(\vec{k})}:=\frac12[\hat{a}^{R}(\vec{k})\hat{a}^{R\dagger}(\vec{k})+\hat{a}^{L}(\vec{k})\hat{a}^{L\dagger}(\vec{k})-1]\,.
\end{split}
\end{equation}
which satisfies the $su(1,1)$ Lie-algebra
\begin{equation}\label{su11Lie}
  [\hat{L}_0^{(\vec{k})},\hat{L}_\pm^{(\vec{k})}]=\pm\hat{L}_\pm^{(\vec{k})}\,, \qquad [\hat{L}_-^{(\vec{k})},\hat{L}_+^{(\vec{k})}]=2\hat{L}_0^{(\vec{k})}\,.
\end{equation}

In general, the tangent operator $\hat{T}(s)$ has the form
\begin{equation}\label{expressT1}
  \hat{T}(s)=\int\td^{d-1}k[\alpha_{+}(s)\hat{L}_+^{(\vec{k})}+\alpha_{-}(s)\hat{L}_-^{(\vec{k})}+\alpha_{0}(s)\hat{L}_0^{(\vec{k})}]\,,
\end{equation}
and we have
\begin{equation}\label{UexpressT1}
  \hat{O}(s)=\overleftarrow{P}\exp\left[\int_{0}^s\hat{T}(s)\td s\right] \,.
\end{equation}
In order to compute the complexity between $|\text{TFD}(t_L,t_R)\rangle$ and $|0\rangle$, we need an $s$-dependent operator $\hat{O}(s)$  satisfying
\begin{equation}\label{requiU1}
   \hat{O}(0)=I, ~~~ \hat{O}(1)|0\rangle=|\text{TFD}(t_L,t_R)\rangle\,.
\end{equation}
Since, for different $s_1$ and $s_2$, the generators $\hat{T}(s_1)$ and $\hat{T}(s_2)$ do not commute, we cannot drop the time order operator $\overleftarrow{P}$ in \eqref{UexpressT1}. However, as the generator set~\eqref{Eset1} forms a complete Lie-algebra, there are three functions $b(\vec{k},s), c(\vec{k},s)$ and $d(\vec{k},s)$ so that the operator $ \hat{O}(s)$ can have a
%%
%%
%\begin{equation}\label{UexpressT1}
%  \hat{U}(s)=\exp\left[\int\td^{d-1}k[\alpha_+(\vec{k},s)\hat{L}_+^{(\vec{k})}+\alpha_-(\vec{k},s)\hat{L}_-^{(\vec{k})}+\alpha_0(\vec{k},s)\hat{L}_0^{(\vec{k})}]\right]
%\end{equation}
%%
%These should satisfy following constraint,
%%
%\begin{equation}\label{reqir1}
%  |\psi(1)\rangle:=\hat{U}(1)|0\rangle=|\text{TFD}(t_L,t_R)\rangle\,.
%\end{equation}
%%
%The operator defined in Eq.~\eqref{UexpressT1} can have
 ``normal decomposition'' by using the decomposition formula of the su(1,1) Lie-algebra~\cite{Chapman:2017rqy,klimov2009a}
\begin{equation}\label{decomp11}
\begin{split}
  \hat{O}(s)=&\exp\left[\int\td^{d-1}kb(\vec{k},s)\hat{L}_+^{(\vec{k})}\right]\exp\left[\int\td^{d-1}kc(\vec{k},s)\hat{L}_0^{(\vec{k})}]\right]\times\\
  &\exp\left[\int\td^{d-1}kd(\vec{k},s)\hat{L}_-^{(\vec{k})}\right] \,.
  \end{split}
\end{equation}
%
%In order to use Fubini-Study metric to compute the complexity, we need to find the constraint on $\alpha$ and $\beta$ explicitly. To do so, let us refer to

The requirement $\hat{O}(0)=I$ shows that $b(\vec{k},0)=c(\vec{k},0)=d(\vec{k},0)=0$. One important point of the decomposed form \eqref{decomp11} is that
\begin{equation}\label{sampleU1}
  |\psi(s)\rangle=\hat{O}(s)|0\rangle=\mathcal{N}(s)\exp\left[\int\td^{d-1}k b(\vec{k},s)\hat{L}_+^{(\vec{k})}\right]|0\rangle\,,
\end{equation}
where $\mathcal{N}(s)$ is a normalization constant factor. The constraint Eq.~\eqref{requiU1} with Eqs. \eqref{timeTFD5b} and  \eqref{funcfak} yield
\begin{equation}\label{constaint1}
b_+(\vec{k},1)=e^{-\omega_{\vec{k}}[\pi/a+i(t_R+t_L)]}=\tanh r_{\vec{k}}\,,
\end{equation}
where $r_{\vec{k}}$ is defined by Eq.~\eqref{defrk}. Following Ref.~\cite{Chapman:2017rqy}, we can find the complexity between $|\text{TFD}(t_L,t_R)\rangle$ and $|0\rangle$
\begin{equation}\label{FSc1}
  \mathcal{C}(|\text{TFD}(t_L,t_R)\rangle,|0\rangle)=\min\left\{\frac{\Sigma_{d-1}}2\int_0^1\td s\int\td^{d-1}k\frac{|\partial_s b (\vec{k},s)|}{1-|b(\vec{k},s)|^2}\right\}
\end{equation}
with the constraint Eq.~\eqref{constaint1}. The solution for this optimization problem has been shown \cite{Chapman:2017rqy}:
\begin{equation}\label{solutiongamma1}
  b(\vec{k},s)=\tanh(r_{\vec{k}}s)=\tanh\{s\cdot\text{arctanh}e^{-\omega_{\vec{k}}[\pi/a+i(t_R+t_L)]}\}
\end{equation}
and the complexity is given by,
\begin{equation}\label{FSc2}
  \Sigma_{d-1}^{-1}\mathcal{C}(|\text{TFD}(t_L,t_R)\rangle,|0\rangle)=\int\td^{d-1}k|r_{\vec{k}}|=\int\td^{d-1}k \left|\text{arctanh}e^{-\omega_{\vec{k}}[\pi/a+i(t_R+t_L)]}\right| \,.
\end{equation}

For the full conformal symmetry case, we have $\omega_{\vec{k}}=k$, then Eq.~\eqref{FSc2} becomes
\begin{equation}\label{FSc3}
\begin{split}
  \Sigma_{d-1}^{-1}\mathcal{C}(|\text{TFD}(t_L,t_R)\rangle,|0\rangle)&=S_{d-2}\int_0^\infty\td k k^{d-2}\left|\text{arctanh}e^{-k[\pi/a+i(t_R+t_L)]}\right|\\
  &=2^{d-1}S_{d-2}T^{d-1}\Xi_d(\tilde{t})\,,
  \end{split}
\end{equation}
where $S_{d-2}$ is the area of $(d-2)$-dimensional sphere and $\tilde{t}:=2(t_L+t_R)T$. $\Xi_d(\tilde{t})$ is a function defined as ($x := k/2T$)
\begin{equation}\label{defineRdt}
  \Xi_d(\tilde{t}):=\int_0^\infty x^{d-2}\left|\text{arctanh}e^{-(1+i \tilde{t})x}\right|\td x\,,
\end{equation}
which is finite only when $d\geq2$.
It is more convenient to write the result in terms of the total energy of the system.  For the free scalar field with conformal symmetry, the total energy $E$ is expressed by
\begin{equation}\label{totalECFT}
  \frac{E}{\hbar\Sigma_{d-1}}=\int \td k^{d-1}\omega_{\vec{k}}e^{-\omega_{\vec{k}}/(2T)}=S_{d-2}2^dT^d\int_0^\infty x^{d-1}e^{-x}\td x=S_{d-2}2^d\Gamma(d+1)T^d\,,
\end{equation}
so we have
\begin{equation}\label{FSintermE}
  \mathcal{C}(|\text{TFD}(t_L,t_R)\rangle,|0\rangle)=\frac{\Xi_d(\tilde{t})}{2\Gamma(d+1)}\frac{E}{\hbar T}\,.
\end{equation}

\begin{figure}
  \centering
  \includegraphics[width=.49\textwidth]{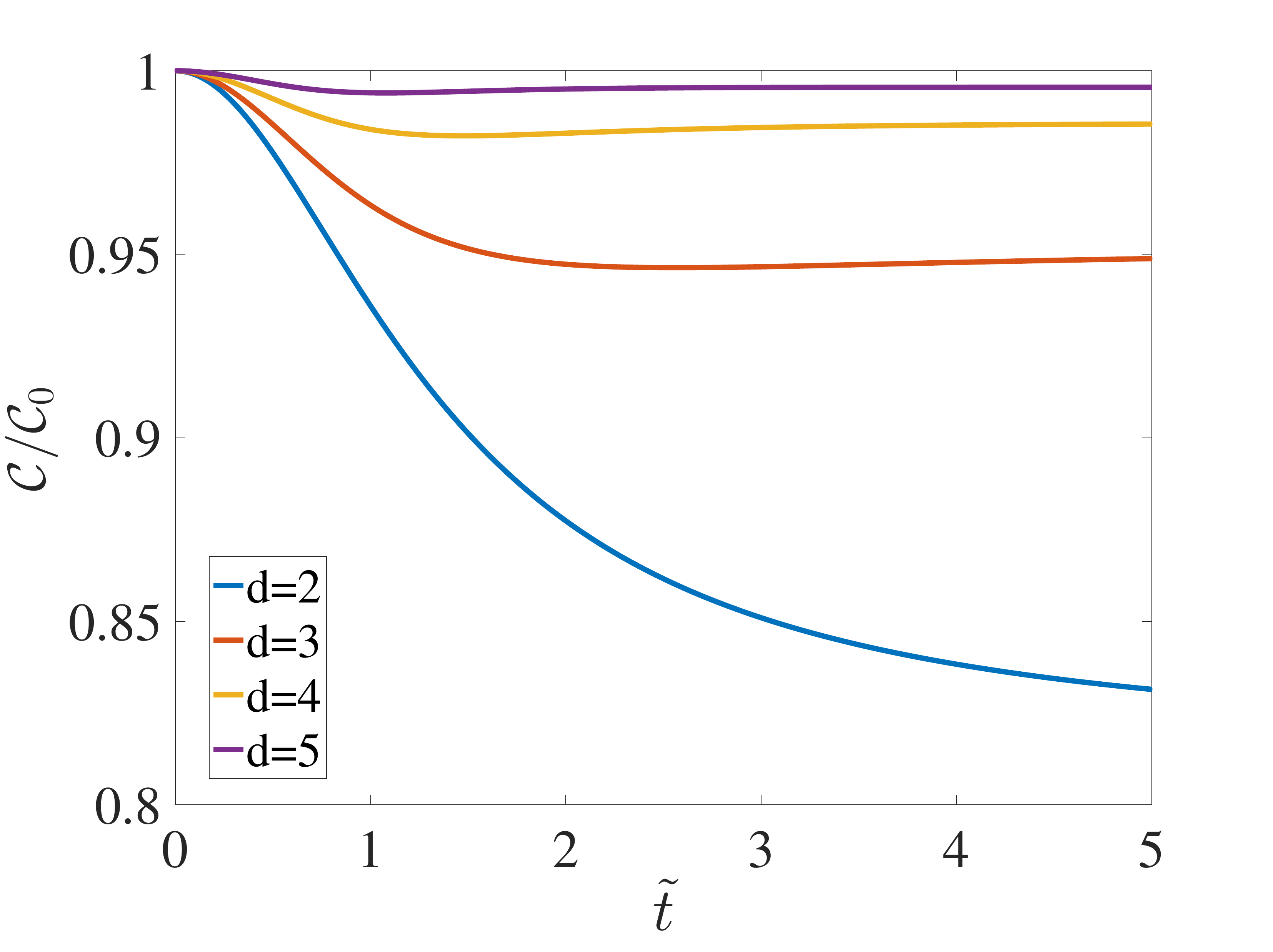}
  \includegraphics[width=.49\textwidth]{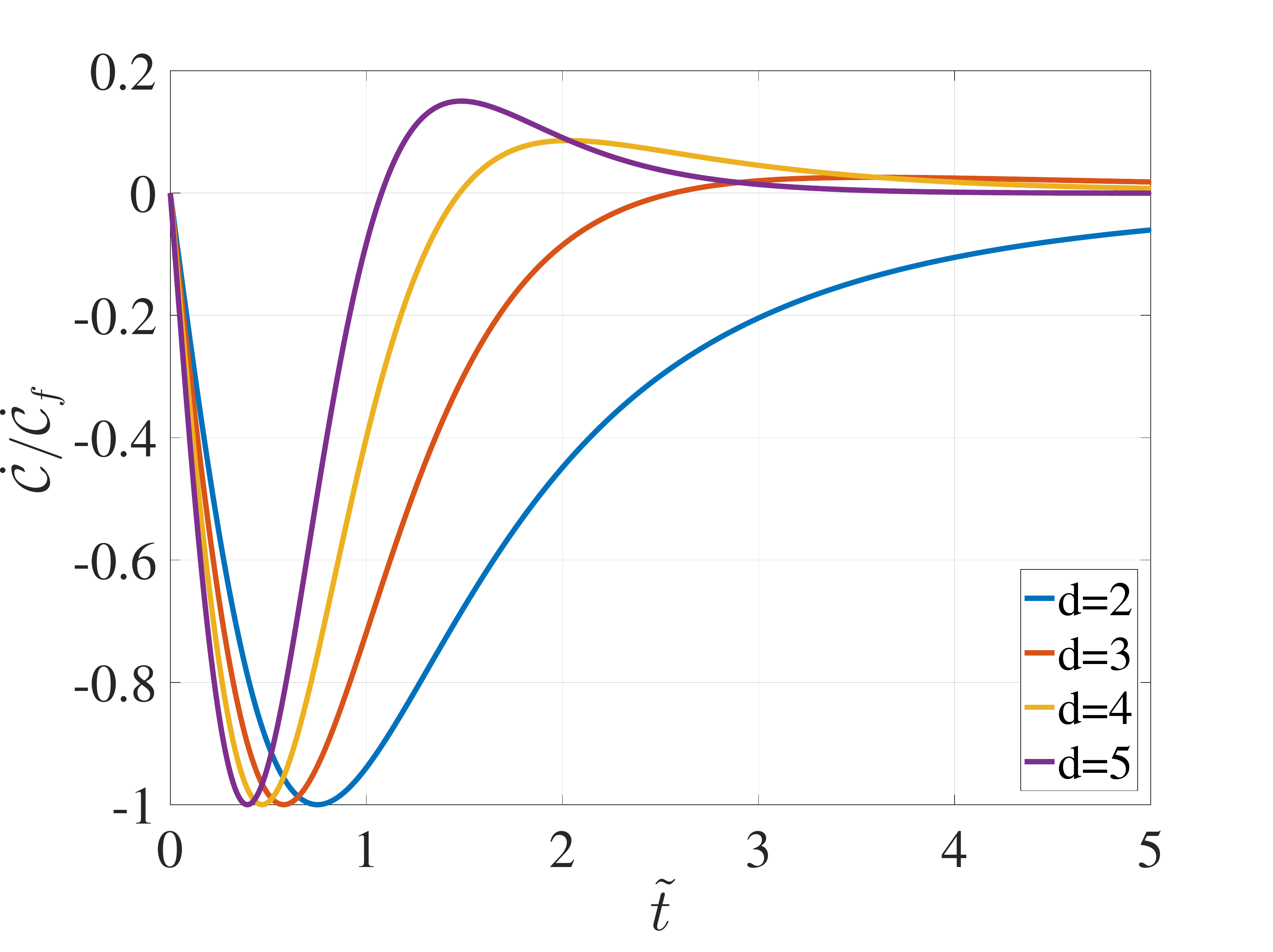}
  \caption{The numerical values of $\mathcal{C}$ and $\dot{\mathcal{C}}$ at $d=2,3,4,5$. Here $\mathcal{C}_0$ is the complexity when $\tilde{t}=0$ and $\dot{\mathcal{C}}_f=-\dot{\mathcal{C}}_{\min}$. They show that $\dot{\mathcal{C}}$ will first decease linearly with respective to time $\tilde{t}$ and then increase later. Finally, the $\dot{\mathcal{C}}$ goes to zero. } \label{Xit1}
\end{figure}

The growth rate of the complexity between $|\text{TFD}(t_L,t_R)\rangle$ and $|0\rangle$ can be expressed as
\begin{equation}\label{growthrate1}
  \frac{\td}{\td(t_L+t_R)}\mathcal{C}(|\text{TFD}(t_L,t_R)\rangle,|0\rangle)=\frac{E}{\hbar \Gamma(d+1)}\dot{\Xi}_d(\tilde{t}) \,.
\end{equation}
For small time $\tilde{t}$ we have the following expansion
\begin{equation}\label{smalltR}
  \Xi_d(\tilde{t})=\mathfrak{I}_d^{(0)}-\frac12\mathfrak{I}_d^{(1)}\tilde{t}^2+\mathcal{O}(\tilde{t}^4)\,,
\end{equation}
where $\mathfrak{I}_d^{(0)}=\Gamma(d)(2^d-1)\zeta(d)/[2^d(d-1)]$ and $\mathfrak{I}_d^{(1)}>0$. It is not easy to write down the  analytic formula for $\mathfrak{I}_d{(1)}$ so the numerical computations shows that
\begin{equation}
\mathfrak{I}_3^{(1)}\approx0.07565\,, \qquad \mathfrak{I}_4^{(1)}\approx0.1639.
\end{equation}
For large $\tilde{t}$ limit, i.e., in the late time limit, we can see that the phase factor $ik\tilde{t}$ makes a rapidly oscillation so the complexity becomes constant
\begin{equation}\label{largetR}
  \lim_{t\rightarrow\infty}\Xi_d(\tilde{t})=\mathfrak{I}_d^{(0)}\vartheta \,,
\end{equation}
with a positive constant $\vartheta\approx0.986$.
Thus we conclude
\begin{equation}\label{limitvalues1}
  \frac{\td}{\td(t_L+t_R)}\mathcal{C}(|\text{TFD}(t_L,t_R)\rangle,|0\rangle)=\left\{
  \begin{split}
  &-\frac{E}{\hbar\Gamma(d+1)}\mathfrak{I}_d^{(1)}\tilde{t} ~~ \quad \text{for}~|\tilde{t}|\ll1,\\
  &\qquad \qquad 0 \qquad  \qquad \ \, \quad  \text{for}~|\tilde{t}|\rightarrow\infty.
  \end{split}\right.
\end{equation}
In Fig.~\ref{Xit1}, the values of $\mathcal{C}(|\text{TFD}(t_L,t_R)$ and its growth rates for different $\tilde{t}$ are shown.
We see that in the Fubini-Study metric method, the complexity growth rate between a TFD state and its corresponding vacuum state is  negative for small time and increases later, finally goes to zero in the large time limit. This is very different from the CV and CA conjectures and also the FS method, which will be considered in the following subsection.

\subsection{Finsler geometry (FG)}
In this subsection, we will use another field theoretic method proposed by Ref.~\cite{Yang:2017nfn} to compute the time dependence of the complexity in the TFD state. Ref.~\cite{Yang:2017nfn} first try to define the complexity for an operator and then define the complexity between two states. Let us first make a brief review on this method.

For a given generator set  $E=\{M^1,M^2,\cdots\}$,  all the operators generated by (Eqs.~\eqref{mapus1}  and \eqref{decompT})
\begin{equation}\label{mapus122}
  \hat{O}(s):=\overleftarrow{P}\exp\left[\int_{0}^s\hat{T}(\tilde{s})\td \tilde{s}\right]\,, \qquad   \hat{T}(s)=\sum_IY_I(s)M^I\,,
\end{equation}
form an operator set $\mathcal{O}$ where the identity operator $I$ is also included.  Eq.~\eqref{mapus122}  defines a curve, $c(s)$, in $\mathcal{O}$, $c: [0,1]\mapsto\mathcal{U}$.  The length of the curve may be defined as
\begin{equation}\label{lengthLc}
  L[c]:=\int_0^1\td s F[c(s),\hat{T}(s)]\,,
\end{equation}
where  the Finsler structure $F[c(s),\hat{T}(s)]$ is always positive and some functional of $Y_I(s)$, which depends on the choice of the generators $\hat{T}(s)$ of the curve $c(s)$. The explicit form of the Finsler structure will be explained later on.

Once the length is defined, the complexity of any operator $\hat{O}$ belonging to $\mathcal{O}$ is given by the minimal length from the identity:
\begin{equation}\label{complU1}
  \mathcal{C}(\hat{O}):=\min\{L[c]~|~\forall c: [0,1]\mapsto\mathcal{O}, \ s.t., \ c(0)=I~\text{and}~\exists\lambda\neq0,~c(1)=\lambda\hat{O}\}\,.
\end{equation}
After defining the complexity of an operator, we may define the complexity from one state to another state as
\begin{equation}\label{FCompstates}
  \mathcal{C}(|\psi_2\rangle,|\psi_1\rangle):=\min\{\mathcal{C}(\hat{O})~|~\forall\hat{O}\in\mathcal{O}, s.t., \hat{O}|\psi_1\rangle\sim|\psi_2\rangle\}\,,
\end{equation}
where the notation $\sim$ means that two state can be different by a nonzero complex constant. Thus, there are three steps to find the complexity between two states. Firstly, we have to find the complexity of all operators in $\mathcal{O}$. Then, we have to find all the operators which can change the reference state to the target state. Finally, we need to find the minimal complexity of these operators.  For some cases where we only care about the complexity between states it is not necessary to compute the complexity of all the operators. Instead, we can directly solve the following optimization problem
\begin{equation}\label{FCompstates2}
  \mathcal{C}(|\psi_2\rangle,|\psi_1\rangle):=\min\{L[c]~|~\forall c:[0,1] \rightarrow \mathcal{O}, s.t., \hat{O}(s)=c(s),~\hat{O}(0)=I~\text{and}~\hat{O}(1)|\psi_1\rangle\sim|\psi_2\rangle\}\,.
\end{equation}
In some cases, this optimization problem is easier to handle than finding complexity of operators.

In Ref.~\cite{Yang:2017nfn}, a very general generator set formed by creation and annihilation operators is considered. Although this makes the generator set big enough, it makes the optimization problems~\eqref{complU1} and \eqref{FCompstates} hard to solve exactly. In order to make a comparison with the results in the Fubini-Study metric, we will use a smaller generator set, which is  defined by~\eqref{Eset1}. In this case, the operator set $\mathcal{O}$ is just the infinite direct product of $SU(1,1)$ group.
Any operator in $\mathcal{O}$ can be parameterized uniquely by three complex-valued functions $\gamma_+(\vec{k}),\gamma_-(\vec{k})$ and $\gamma_0(\vec{k})$.
\begin{equation}\label{decomp1}
\begin{split}
 \hat{U}[\gamma_+(\vec{k}),\gamma_-(\vec{k}),\gamma_0(\vec{k})]=&\exp\left[\int\td^{d-1}k\gamma_+(\vec{k})\hat{L}_+^{(\vec{k})}\right]\exp\left[\int\td^{d-1}k\ln\gamma_0(\vec{k})\hat{L}_0^{(\vec{k})}]\right]\times\\
  &\exp\left[\int\td^{d-1}k\gamma_-(\vec{k})\hat{L}_-^{(\vec{k})}\right]\,.
  \end{split}
\end{equation}
%
%The value of $\{\gamma_\pm(\vec{k}), \gamma_0(\vec{k})\}$ in terms of $\{\alpha_\pm(\vec{k}), \alpha_0(\vec{k})\}$ can be expressed as,
%%
%\begin{equation}\label{value3gamma}
%  \Lambda^2:=\frac{\alpha_0^2}4-\alpha_+\alpha_-,~~\gamma_0=\left(\cosh\Lambda-\frac{\alpha_0}{2\Lambda}\sinh\Lambda\right)^{-2},~~\gamma_\pm=\frac{2\alpha_\pm\sinh\Lambda}{2\Lambda\cosh\Lambda-\alpha_0\sinh\Lambda}\,.
%\end{equation}
%%
We find from \eqref{funcfak} that $|\text{TFD}(t_L,t_R)\rangle\sim \hat{U}[\gamma_+(\vec{k}),\gamma_-(\vec{k}),\gamma_0(\vec{k})]|0\rangle$ if and only if $\gamma_+=e^{-\omega_{\vec{k}}[\pi/a+i(t_R+t_L)]}$. Let us take
\begin{equation}\label{Ugamma}
  \hat{U}_a(t_L,t_R):=\exp\left[\int\td^{d-1}ke^{-\omega_{\vec{k}}[\pi/a+i(t_R+t_L)]}\hat{L}_+^{(\vec{k})}\right]\,.
\end{equation}
Thus the set of all the operators which can change from $|0\rangle$ to $|\text{TFD}(t_L,t_R)\rangle$ is
\begin{equation}\label{goodset}
  \mathcal{D}:=\left\{\left.\hat{U}[\gamma_+(\vec{k}),\gamma_-(\vec{k}),\gamma_0(\vec{k})]~\right|~\forall~\gamma_-,\gamma_0\in\mathbb{C}, \gamma_+=e^{-\omega_{\vec{k}}[\pi/a+i(t_R+t_L)]}\right\} \,.
\end{equation}
The complexity between $|0\rangle$ to $|\text{TFD}(t_L,t_R)\rangle$ is given by
\begin{equation}\label{FGTFDA2}
\begin{split}
  \mathcal{C}(|\text{TFD}(t_L,t_R)\rangle,&|0\rangle)=\min\left\{\mathcal{C}(\hat{U})|~\forall \hat{U}\in\mathcal{D}\right\}
  \end{split}
\end{equation}

In order to proceed, we need to obtain the explicit form of the Finsler structure in the generator set $E_L$.
Relegating technical details to appendix~\ref{FinSEL} we here present a final result.
For any generator $\hat{T}(s)$ expanded in the basis $E_L$
\begin{equation}\label{FGgenerator0tt}
  \hat{T}(s)=\int\td^{d-1}k[\alpha_+(s,\vec{k})\hat{L}_+^{(\vec{k})}+\alpha_0(s,\vec{k})\hat{L}_0^{(\vec{k})} +\alpha_-(s,\vec{k})\hat{L}_-^{(\vec{k})}]\,.
\end{equation}
the Finsler structure is given by
\begin{equation}\label{defFp20tt}
  F|_{E_L}=\ell_0\Sigma_{d-1}\int\td^{d-1}k[\parallel\alpha_+(s,\vec{k})\parallel+\parallel\alpha_-(s,\vec{k})\parallel +\parallel\alpha_0(s,\vec{k})\parallel]\,,
\end{equation}
where $\ell_0$ is a free parameter to be chosen later. Based on the detailed computation in appendix~\ref{relga}, it turns out that
\begin{equation}\label{FGTFDA4tt}
 \mathcal{C}(|\text{TFD}(t_L,t_R)\rangle,|0\rangle)=\ell_0 \Sigma_{d-1} \int\td^{d-1}k\parallel\gamma_+(\vec{k}) \parallel \,.
\end{equation}
Therefore, we have
\begin{equation}\label{valueFGC2}
  \mathcal{C}(|\text{TFD}(t_L,t_R)\rangle,|0\rangle)  = \ell_0\Sigma_{d-1}\int_0^\infty k^{d-2}\parallel e^{-\omega_{\vec{k}}[\pi/a+i(t_R+t_L)]}\parallel\td k\,.
\end{equation} 
Using the definition $\parallel\cdot\parallel$ in Eq.~\eqref{complexY}, we finally obtain the complexity between $|0\rangle$ to $|\text{TFD}(t_L,t_R)\rangle$:
\begin{equation}\label{valueFGC3}
  \mathcal{C}(|\text{TFD}(t_L,t_R)\rangle,|0\rangle)=\ell_0\Sigma_{d-1}S_{d-2}2^{d-1}T^{d-1}\Omega_d(\tilde{t})=\ell_0\frac{\Omega_d(\tilde{t})}{2\Gamma(d+1)}\frac{E}{\hbar T}\,,
\end{equation}
where $S_{d-2}$ is the area of $(d-2)$-dimensional sphere, $\tilde{t}:=2(t_L+t_R)T$,
\begin{equation}\label{defOmegat}
  \Omega_d(\tilde{t}):=\int_0^\infty x^{d-2}e^{-x}(|\cos x \tilde{t}|+|x \tilde{t}|\cdot|\sin x \tilde{t}|)\td x\,,
\end{equation}
and the total energy $E$ is given by Eq.~\eqref{totalECFT}.

\begin{figure}
  \centering
  \includegraphics[width=.49\textwidth]{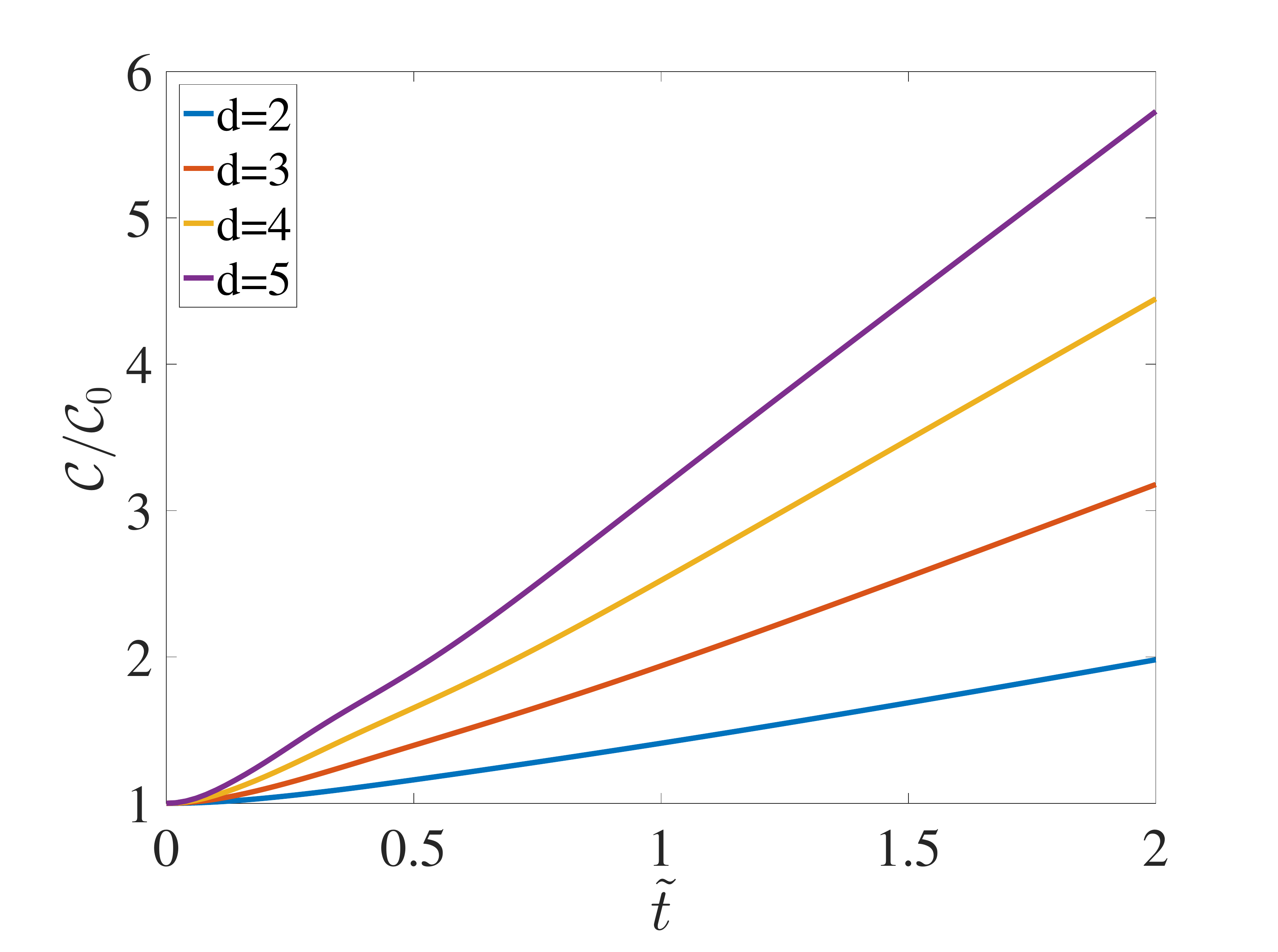}
  \includegraphics[width=.49\textwidth]{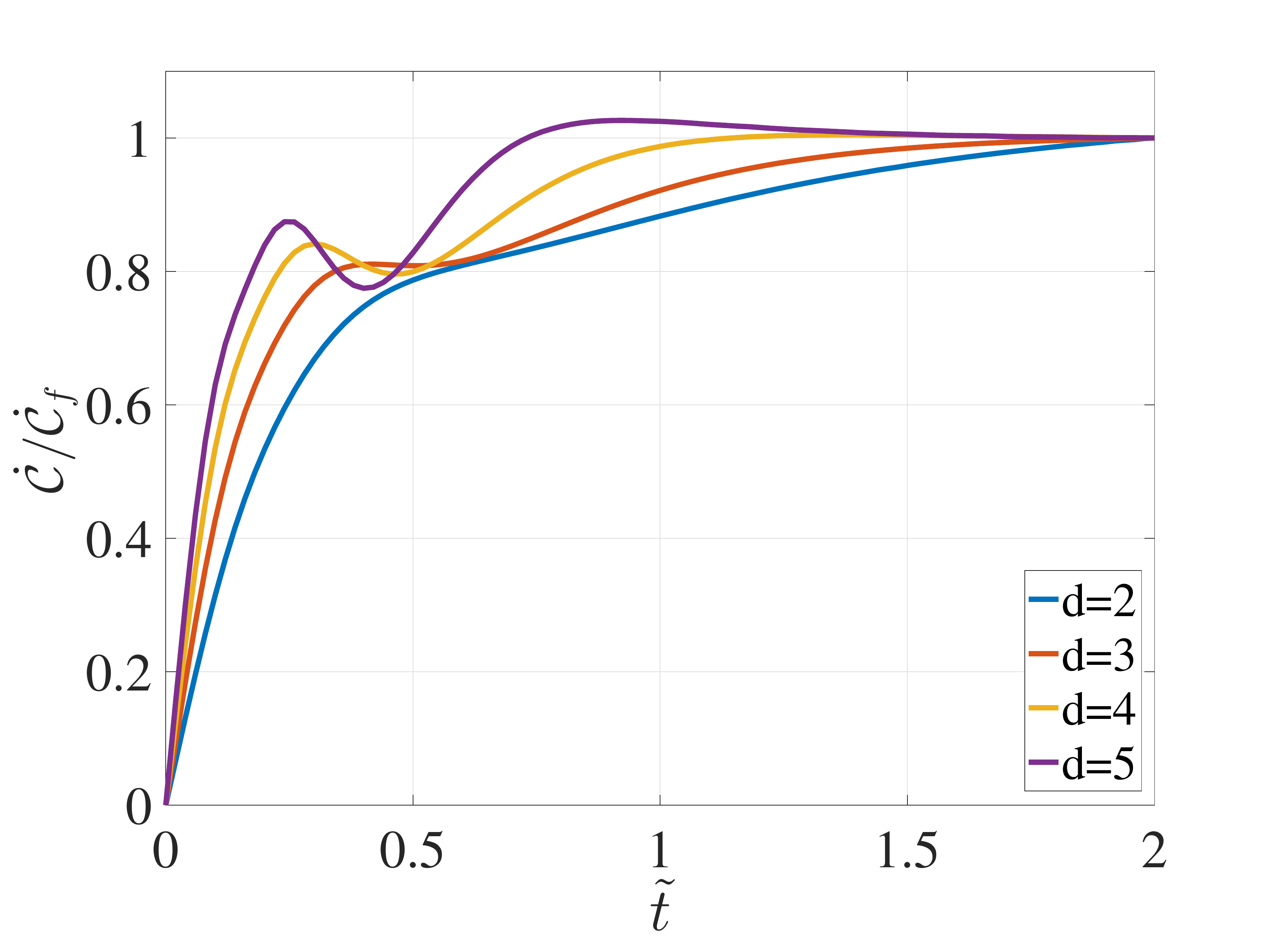}
  \caption{The numerical values of complexity and its growth rate at $d=2,3,4,5$. $\mathcal{C}_0$ is the complexity at $\tilde{t}=0$ and $\dot{\mathcal{C}}_f$ is the growth rate at the late time limit. From the left panel, we see that $\mathcal{C}$ will monotonously increase with respective to time $\tilde{t}$.  For small $\tilde{t}$, $\dot{\mathcal{C}}$ will linearly depends on $\tilde{t}$. For large $\tilde{t}$, $\dot{\mathcal{C}}$ will tend to increase linearly with respective to $\tilde{t}$ and $\dot{\mathcal{C}}$ tends to a constant. For large $d$, we obtain similar behaviours. } \label{valuesW}
\end{figure}

For small $\tilde{t}$
\begin{equation}\label{smalltOmega}
  \Omega_d(\tilde{t})=\Gamma(d-1)+\frac{\Gamma(d+1)}2\tilde{t}^2+\mathcal{O}(\tilde{t}^4)\,,
\end{equation}
and for large $\tilde{t}$
\begin{equation}\label{largetOmega}
  \Omega_d(\tilde{t})=\frac2{\pi}\left[\Gamma(d-1)+\Gamma(d)\tilde{t}\right][1+\mathcal{O}(1/\tilde{t})]\,.
\end{equation}
Thus we see that
\begin{equation}\label{limitvalues3}
  \frac{\td}{\td(t_L+t_R)}\mathcal{C}(|\text{TFD}(t_L,t_R)\rangle,|0\rangle)=\left\{
  \begin{split}
  &\ell_0E\tilde{t}/\hbar~~~~\text{for}~|\tilde{t}|\ll1,\\
  &\ell_0\frac{E}{\hbar d}~~~~~~~\text{for}~|\tilde{t}|\gg1.
  \end{split}\right.
\end{equation}
For the intermediate time, we can compute $\Omega_d(\tilde{t})$ analytically but it is not so illuminating. Therefore, we show
a numerical plot for $\Omega_4(\tilde{t})$ and $\dot{\Omega}_4(\tilde{t})$ in the Fig.~\ref{valuesW}. For lager $d>4$, the behavior is similar.

Note that the linear $\tilde{t}$ dependence of the complexity in the late time limit comes from $|x \tilde{t}|$ in Eq. \eqref{defOmegat}, which is due to our definition of  $\parallel\cdot\parallel$ in  Eq.~\eqref{complexY}.  If we use the definition  $\parallel\cdot\parallel$ in Eq.~\eqref{complexY0} then the complexity will be constant independent of time, which is the same as the FS case. Therefore, our result here is not so robust. It should be understood as one example to define the field theory complexity showing the linear-time complexity.

The complexity growth rate is positive and linearly dependent of $t$ at the early time ($(t_L+t_R)T\ll1$), which is  the same as the prediction of the CV conjecture. In the late time limit ($(t_L+t_R)T\gg1$), Eq.~\eqref{limitvalues3} is constant and proportional to $T^{d}$. In the planar symmetry asymptotic AdS black hole, the total ADM mass $M$ is also proportional to $T^{d}$. Thus we see that the complexity growth rate is similar to the predictions of both the CV and CA conjectures in the  late time limit. The free parameter $\ell_0$ in Eq.~\eqref{valueFGC2} can be determined if we require that the complexity growth rate  saturate to the Lloyd's bound at $t_L+t_R\rightarrow\infty$. We see that if we take
$$\ell_0=\frac{2d}\pi\,,$$
the complexity growth rate at the late time limit saturates to the Lloyd's bound. However,
it turned out that the subleading term in Eq.~\eqref{largetOmega} is positive so the complexity growth rate will approach to the limiting value from the larger value. Thus, like the CA conjecture, at large time region, the complexity growth rate violate the Lloyd's bound with this choice of $\ell_0$.

\section{Summary}\label{summ}

In this paper, we have computed the complexity of the time dependent TFD states and their growth rates by four different methods, two holographic and two field theory methods. Two holographic methods are based on the ``complexity-action'' (CA) conjecture or ``complexity-volume'' (CV) conjecture. Two quantum field theoretic methods are based on the Fubini-Study metric (FS) or the Finsler geometry (FG). In particular, for holographic computation, we have proposed a modified CA and CV conjectures between two TFD states, $|\text{TFD}_2\rangle$ and $|\text{TFD}_1\rangle$\eqref{newcvca}
\begin{equation}
\begin{split}
 &\mathcal{C}_V(|\text{TFD}_2\rangle,|\text{TFD}_1\rangle)\equiv |\mathcal{C}^{(1)}_{V}-\mathcal{C}^{(2)}_{V}|\,, \\
 &\mathcal{C}_A(|\text{TFD}_2\rangle,|\text{TFD}_1\rangle)\equiv |\mathcal{C}^{(1)}_{A}-\mathcal{C}^{(2)}_{A}|\,,
 \end{split}
\end{equation}
where ${\mathcal{C}}_{V}^{(i)}$ and ${\mathcal{C}}_{A}^{(i)}$ are the original CV and CA conjectures for the $|\text{TFD}_i\rangle$ state. It is similar to the `complexity of formation' proposed in \cite{Chapman:2016hwi} but there is a subtle difference in that here we do not assume any reference state~\cite{Yang:2017nfn}. These modified versions yield finite values agreeing to the field theory computation for a static case~\cite{Yang:2017nfn}.
For a concrete example in this paper we consider the complexity between the time-dependent TFD state and its corresponding vacuum state. We call it `complexity' for simplicity.

\begin{figure}
  \centering
  \includegraphics[width=.49\textwidth]{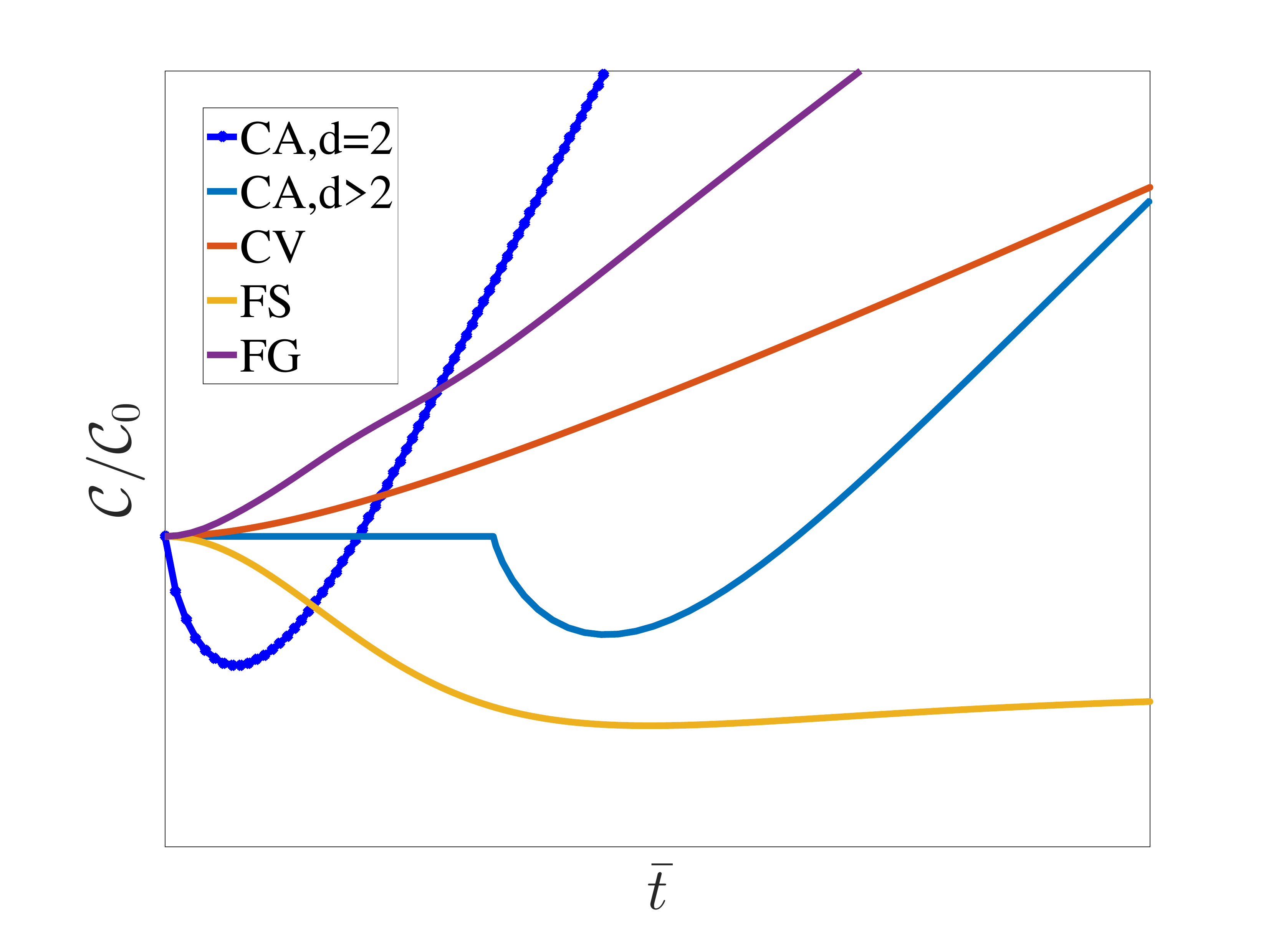}
  \includegraphics[width=.49\textwidth]{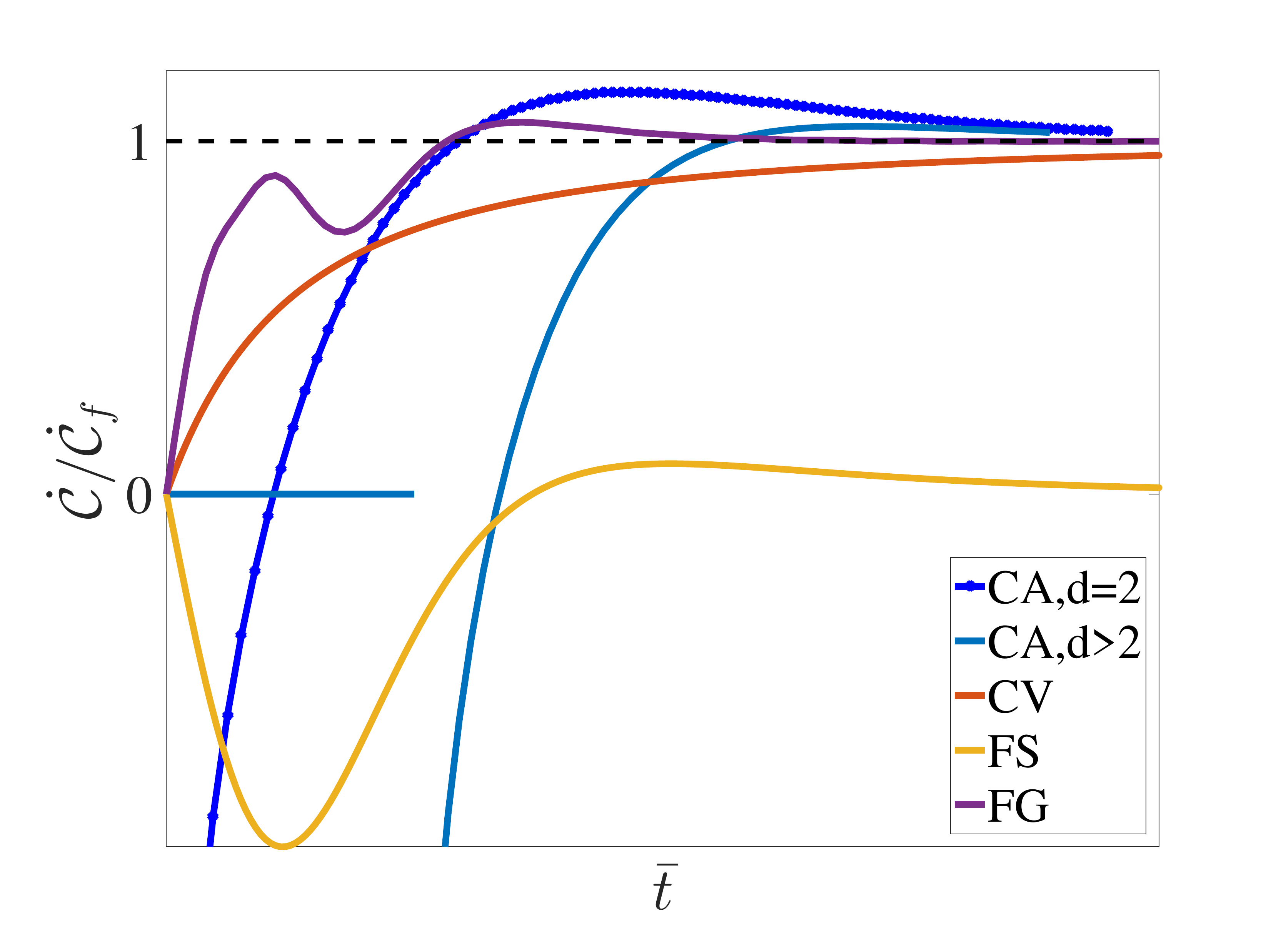}
  \caption{Schematic plots for the complexity by four methods: the holographic CV and CA conjecture and the field theoretic FS and FG methods. It is obtained from Figures \ref{FigAdS1}, \ref{FigBTZ1}, \ref{Figrcv}, \ref{Xit1}, and \ref{valuesW}.  $\bar{t}=  t_L + t_R$. Left: ${\mathcal{C}}/{\mathcal{C}}_0$, where ${\mathcal{C}}_0$ is the complexity at $\bar{t}=0$.  Right: $\dot{\mathcal{C}}/\dot{\mathcal{C}}_f$, where $\dot{\mathcal{C}}_f$ is the growth rate at $\bar{t} \rightarrow\infty$ in CA,CV and FG methods, and $\dot{\mathcal{C}}_f=-\dot{\mathcal{C}}_{\min}$ in the FS method.}
 \label{summaryfig}
\end{figure}

Our main results for the time dependent complexity for the TFD states are summarized in Table~\ref{tab1}. As a companion to Table~\ref{tab1}, for readers' convenience, we show a schematic plot, Fig. \ref{summaryfig} of which precise information can be found in Figures \ref{FigAdS1}, \ref{FigBTZ1}, \ref{Figrcv}, \ref{Xit1}, and \ref{valuesW}. We define a common time $\bar{t} = t_L+t_R$ for all cases.

\begin{table}
\centering
\begin{tabular}{c|c|c|c|c}
  \hline
  ~ & CA & CV & FS & FG \\
   \hline
   $t_L+t_R=0$ & $\mathcal{C}\propto\frac{E}{\hbar T}$ &  $\mathcal{C}\propto\frac{E}{\hbar T}$ & $\mathcal{C}\propto\frac{E}{\hbar T}$ &$\mathcal{C}\propto\frac{E}{\hbar T}$\\
   %\hline
  early time & $\dot{\mathcal{C}}=0$ if $d>2$ & $\dot{\mathcal{C}}\propto \bar{t}$ & $\dot{\mathcal{C}}\propto-\bar{t}$ & $\dot{\mathcal{C}}\propto \bar{t}$ \\
  ~ & $\dot{\mathcal{C}}=-\infty$ if $d=2$ & ~& ~ & ~\\
  late time & $\dot{\mathcal{C}}=\frac{2E}{\pi\hbar}$ & $\dot{\mathcal{C}}=\frac{2E}{\pi\hbar}$ & $\dot{\mathcal{C}}=0$ & $\dot{\mathcal{C}}=\frac{2E}{\pi\hbar}$ \\
  sign$(\dot{\mathcal{C}})$& indefinite &+&indefinite &+\\
  Lloyd's bound&broken&satisfied&saisfied&broken\\
  \hline
\end{tabular}
\caption{The summary of the complexity between the time-dependent TFD state and its corresponding vacuum state in four different methods. $\bar{t}=t_L+t_R$ and $E$ is the total energy of the system. We have set $\ell/\el=4\pi^2\hbar/(d-1)$ for the CV conjecture, $\ell_0=2d/\pi$ for the FG method, and the speed of light $c=1$.}\label{tab1}
\end{table}

If $\bar{t} =0$, four different methods give similar results but give different predictions on the time evolution of the complexity.
At early time, both the CV conjecture and FG method predict that the complexity will increase as $\bar{t}^2$ while the FS method predicts that the complexity will decrease as $-\bar{t}^2$.
The CA conjecture says that for $d>2$ the  complexity does not change until a critical time and after then it will decrease.
For $d=2$, it decreases first and increases as time goes on.

In the late time limit, the CA conjecture, CV conjecture and FG method predict that the complexity will increase linearly in  $\bar{t}$ and the growth rate will be proportional to the total energy of the system. On the contrary, the FS method shows that the complexity in the late time limit will keep constant rather than increasing. The CV conjecture and FG method show that the complexity will monotonically increase for $\bar{t}>0$ while the CA conjecture and FS method show that the complexity first decreases and then increases. The Lloyd's bound is satisfied only for the CV conjecture and FS method. The Lloyd's bound is also satisfied for the CA conjecture and FG method in the late time limit, but it is weakly violated in the intermediate time. We have set $\ell/\el=4\pi^2\hbar/(d-1)$ for the CV conjecture, $\ell_0=2d/\pi$ for the FG method. With other choices, the growth rate saturate to some value which is not the Lloyd's bound.

The results summarized in Table~\ref{tab1} seem to give us some pieces of information to judge which are appropriate methods to compute the complexity among two holographic conjectures and two quantum field theory proposals. For examples, if we expect that the complexity should increase with time then it seems that the CV conjecture and FG method are favored. The similarity between the CV conjecture and the FG method in the early and late time limit seems to show these two proposal are more correlated while they are different in the intermediate time regime.  Note that the FS method is quite different from all the other methods.  In particular, the FS method shows the complexity will keep constant in the late time limit. Because the linear growth of the complexity in the late time limit has much evidence both in quantum information theory and holography~\cite{Susskind:2014moa,Susskind:2014rva,Stanford:2014jda,Brown:2015lvg,Brown:2017jil,Hashimoto:2017fga,Qaemmaqami:2017lzs} it seems to be a challenge to the FS method.

However, there is also a caveat in the FG method. The results of the FG method depend on the definition of the Finsler structure. Our result here should be understood as just one example to define the Finsler structure displaying the linear-time complexity in the late time limit and showing similar behaviors to holographic complexity. For both the FS and FG method, the complexity also depend on the generator set. We have chosen a small generator set to make the TFD states so that we can compute the complexity analytically. If we choose another generator set the complexity may or may not change. Therefore, it will be interesting to investigate how much our results are robust under different choices of generator sets and/or different choices of the Finsler structures.

\acknowledgments
The work of K.-Y. Kim and C. Niu was supported by Basic Science Research Program through the National Research Foundation of Korea(NRF) funded by the Ministry of Science, ICT $\&$ Future Planning(NRF- 2017R1A2B4004810) and GIST Research Institute(GRI) grant funded by the GIST in 2017. C.Y. Zhang is supported by National Postdoctoral Program for Innovative Talents BX201600005.
We also would like to thank the APCTP(Asia-Pacific Center for Theoretical Physics) focus program, ``Geometry and Holography for Quantum Criticality'' in Pohang, Korea for the hospitality during our visit,
where part of this work was done.

\appendix
\section{Fubini-Study metric}\label{FSmetric}
Let us consider an $n$-dimensional Hilbert space $\mathcal{H}$. Any two vector $|\psi_1\rangle$ and $|\psi_2\rangle$ describe the same state if there is a nonzero complex number $c$ such that $|\psi_1\rangle=c|\psi_2\rangle$. This means that the different states of the Hilbert space $\mathcal{H}$ form a complex projective space $\mathbb{C}$P$^n$. As $\mathbb{C}$P$^n=S^{2n+1}/S^1$, we can use the length of geodesic curve in $S^{2n+1}/S^1$ to build the distance of two states. It turns out that this distance is the Fubini-Study distance, which is
\begin{equation}\label{FSdist1}
  D_{FS}(|\psi\rangle,|\phi\rangle):=\arccos(|\langle\psi|\phi\rangle|)\in[0,\pi/2]\,.
\end{equation}
In this equation, the state vectors should satisfy $\langle\psi|\psi\rangle=\langle\phi|\phi\rangle=1$. This expression can be  generalized to  the infinite dimensional cases. To obtain the line element in a local form, let us  assume
\begin{equation}\label{dpsi1}
  |\phi\rangle= \mathcal{N}\left( |\psi\rangle+\td|\psi\rangle \right)
\end{equation}
and expand to second order in the vector $\td|\psi\rangle$.  $\mathcal{N}$ is the normalization factor for $\langle\phi|\phi\rangle=1$. The result is the Fubini-Study metric
\begin{equation}\label{FSmetric1}
 \td s_{FS}^2(|\psi\rangle)=\langle\td\psi|\td\psi\rangle^2-|\langle\td\psi|\psi\rangle|^2\,.
\end{equation}
For any curve $l:[s_i,s_f]\mapsto\mathcal{H}$ such that $l(t)=|\psi(t)\rangle$, its length is defined by
\begin{equation}\label{lengthl}
  L^2[l]:=\int_{s_i}^{s_f}\td s_{FS}(|\psi(t)\rangle)=\int_{s_i}^{s_f}\td t\sqrt{|\partial_t|\psi(t)\rangle|^2-|\langle\psi(t)|\partial_t|\psi(t)\rangle|^2}
\end{equation}
If there is no restriction for the curve, the length of the geodesic connecting any two states is given by Eq.~\eqref{FSdist1}. We can define the complexity for two states
\begin{equation}\label{defC1}
    \mathcal{C}(|\psi\rangle,|\phi\rangle):=D_{FS}(|\psi\rangle,|\phi\rangle)=\arccos(|\langle\psi|\phi\rangle|)\in[0,\pi/2]\,.
\end{equation}
However, in the case that the curve can only be generated by some appointed generator set $E$, the minimal length of the  curves may be different from the Eq.~\eqref{FSdist1} and we have to solve the following optimization problem
\begin{equation}\label{defC2}
    \mathcal{C}(|\psi\rangle,|\phi\rangle,E):=\min_{E}\{L^2[l]~|~l(s_i)=|\psi\rangle,l(s_f)=|\phi\rangle\}\,.
\end{equation}
As noted by Ref.~\cite{Chapman:2017rqy}, it is not necessary to restrict the line element in $L^2$ normal. Then for more general case, we can define the general Fubini-Study metric by $L^p$ normal, which reads
\begin{equation}\label{lengthl}
  \td s:=[||\partial_s|\psi(s)\rangle|^p-|\langle\psi(s)|\partial_s|\psi(s)\rangle|^p|]^{1/p} \,.
\end{equation}

\section{Finsler structure in the generator set $E_L$}\label{FinSEL}
In this appendix, we  explain how to obtain the explicit functional form of the Finsler structure. We start with  the proposal for a more general case in Ref.~\cite{Yang:2017nfn} and we restrict ourselves to the generator set $E_L$.

In Ref.~~\cite{Yang:2017nfn}, the generator set is chosen as the general enveloping algebra of Heisenberg-Weyl Lie algebra. Let us first define the fundamental generator set $E^{0}$ to be the collection of all the creation and annihilation operators
\begin{equation}\label{defE0}
  E^{0}:=\bigcup_{i}\{\hat{a}_i^\dagger,\hat{a}_i, \hat{\mathbb{I}}\}\,.
\end{equation}
Here index $i$ stands for different creation and annihilation operators (in our context, $i$ may stand for $\vec{k}$ and the superscript $R$ and $L$.).  $\hat{\mathbb{I}}$ satisfies $[\hat{a}_i,\hat{a}_j^\dagger]=\hat{\mathbb{I}}\delta_{ij}$ and $\hat{\mathbb{I}}\hat{e}=\hat{e}$ for $\forall\hat{e}\in E^{0}$. This fundamental operator set forms a Heisenberg-Weyl Lie algebra. Because this generator set is not big enough Ref.~\cite{Yang:2017nfn} extends it to a larger set $E$ by
\begin{equation}\label{defElarge}
  E:=\bigcup_{n}(E^{0})^n,~~~\text{with}~(E^0)^n:=\{\hat{M}^{i_1i_2\cdots i_n}=:\hat{e}_{i_1}\hat{e}_{i_2}\cdots \hat{e}_{i_n}:|\forall \hat{e}_{i_1},\hat{e}_{i_2},\cdots,\hat{e}_{i_n}\in E^0\}\,.
\end{equation}
Here the ``:~:'' stands for the normal ordering, e.g., $:\hat{a}_{i}\hat{a}^\dagger_{j}:=\hat{a}^\dagger_{j}\hat{a}_{i}$ for $\forall i,j$. In the definition~\eqref{defElarge}, $\hat{e}_{i_1}, \hat{e}_{i_2}, \cdots, \hat{e}_{i_n}$ do not need to be different from each others. This extended operator set forms the general enveloping algebra of Heisenberg-Weyl Lie algebra. Any generator $\hat{T}(s)$ can be expand it by basis $E$ as follows:
\begin{equation}\label{decompTs1}
  \hat{T}(s)=T_0(s)\hat{\mathbb{I}}+\sum_i Y_i(s)\hat{M}^i+\sum_{ij}Y_{ij}(s)\hat{M}^{ij}+\cdots+\sum_{i_1i_2\cdots i_n}Y_{i_1i_2\cdots i_n}(s)\hat{M}^{i_1i_2\cdots i_n}+\cdots\,.
\end{equation}
Here $T_0(s), Y_i(s), Y_{ij}(s),\cdots$ are complex numbers, $\hat{M}^i, \hat{M}^{ij},\cdots$ are the elements in $E$ except for $\hat{\mathbb{I}}$. Then the Finsler structure expressed in the basis \eqref{defElarge} is given by~\cite{Yang:2017nfn}
\begin{equation}\label{defFp10}
  F|_E=\ell_s\left[\sum_i \parallel Y_i(t)\parallel +2\sum_{ij}\parallel Y_{ij}(t)\parallel +\cdots+n\sum_{i_1i_2\cdots i_n}\parallel Y_{i_1i_2\cdots i_n}(t)\parallel +\cdots\right]\,.
\end{equation}
Here $\ell_s$ is a dimensionless positive constant. The meaning of $\parallel\cdot\parallel$ will clarified later on. For the continuous index case, the summation in the right-hand of Eq.~\eqref{defFp10} should be replaced by integration. Here the index $E$ is added into $F$  to explicitly show that the right-hand of Eq.~\eqref{defFp10} is valid only when we use the set $E$ to expand $\hat{T}(s)$.

In this paper, the generator set $E_L$ is neither $E$ nor its subsect, we cannot directly use formula~\eqref{defFp10} to obtain the functional form of the Finsler structure in the basis $E_L$ as the functional form of a Finsler structure depends on the basis. However, the generator set $E_L$ is just the linear combinations of some elements in $E$,
\begin{equation}\label{transforEEL}
  \left[\begin{matrix}
\hat{L}_+^{(\vec{k})}\\
\hat{L}_-^{(\vec{k})}\\
\hat{L}_0^{(\vec{k})}
\end{matrix}\right]=\left[\begin{matrix}1,0,0,0,0\\
0,1,0,0,0\\
0,0,\frac12,\frac12,\frac12
\end{matrix}\right]
\left[\begin{matrix}
\hat{a}^{R\dagger}(\vec{k})\hat{a}^{L\dagger}(\vec{k})\\
\hat{a}^{R}(\vec{k})\hat{a}^{L}(\vec{k})\\
\hat{a}^{R\dagger}(\vec{k})\hat{a}^{R}(\vec{k})\\
\hat{a}^{L\dagger}(\vec{k})\hat{a}^{L}(\vec{k})\\
\hat{\mathbb{I}}
\end{matrix}\right]
\end{equation}
Let us consider a generator $\hat{T}(s)$ expanded it in the basis $E_L$
\begin{equation}\label{FGgenerator0}
  \hat{T}(s)=\int\td^{d-1}k[\alpha_+(s,\vec{k})\hat{L}_+^{(\vec{k})}+\alpha_0(s,\vec{k})\hat{L}_0^{(\vec{k})} +\alpha_-(s,\vec{k})\hat{L}_-^{(\vec{k})}]\,.
\end{equation}
Using the basis transformation formula in Eq.~(A.5) of Ref.~\cite{Yang:2017nfn}, we find that the functional form of the Finsler structure is
\begin{equation}\label{defFp20}
\begin{split}
  F|_{E_L}&=\frac{2\ell_s\Sigma_{d-1}}{(2\pi)^{d-1}}\int\td^{d-1}k[\parallel\alpha_+(s,\vec{k})\parallel+\parallel\alpha_-(s,\vec{k})\parallel +\frac12\parallel\alpha_0(s,\vec{k})\parallel+\frac12\parallel\alpha_0(s,\vec{k})\parallel]\\
  &=\ell_0\Sigma_{d-1}\int\td^{d-1}k[\parallel\alpha_+(s,\vec{k})\parallel+\parallel\alpha_-(s,\vec{k})\parallel +\parallel\alpha_0(s,\vec{k})\parallel]\,,
\end{split}
\end{equation}
where we define that $\ell_0=2\ell_s/(2\pi)^{d-1}$.

The notation $\parallel\cdot\parallel$ was introduced in Ref.~\cite{Yang:2017nfn}. Let's explain again why it is used in the Finsler structure~\eqref{defFp10}. For a complex number $Y^I=\rho^I e^{i\theta^I}$ one may want to use $\parallel Y^I\parallel =\rho^I$ but this will lead to an inconsistent: the ``rotation'' caused by $\theta^I$ will change the operator $ \hat{O}(s)$ but it does not change the complexity. One simple modification will be
\begin{equation} \label{complexY0}
\parallel Y^I(s)\parallel = \rho^I(|\cos\theta^I|+|\sin\theta^I|) = |\text{Re}Y^I(s)|+|\text{Im}Y^I(s)| \,.
\end{equation}
Another possibility is~\cite{Yang:2017nfn}
\begin{equation}\label{complexY}
  \parallel Y^I(s)\parallel :=|\text{Re}Y^I(s)|+|\theta^I(s)|\cdot|\text{Im}Y^I(s)|\,.
\end{equation}
where $\theta(0)\in[-\pi,\pi)$ and $\theta(s)$ is continuous for $s\in[0,1]$. In this paper, we choose the \eqref{complexY} as we can see that it can give the linear growth rate of the complexity at the late time limit.

\section{Complexity of operator generated by $E_L$}\label{relga}
In this appendix, we will give the method to compute the complexity for any element in operators set $\mathcal{U}$ which is generated by generator set \eqref{Eset1}. Any operator in $\mathcal{U}$ can be parameterized uniquely by three complex-valued functions $\gamma_+(\vec{k}),\gamma_-(\vec{k})$ and $\gamma_0(\vec{k})$ by Eq.~\eqref{decomp1}.
%%
%\begin{equation}\label{parameU1}
%  \hat{U}[\gamma_+(\vec{k}),\gamma_-(\vec{k}),\gamma_0(\vec{k})]:=\exp\left\{\int\td^{d-1}k[\alpha_+(\vec{k})\hat{L}_+^{(\vec{k})}+\alpha_-(\vec{k})\hat{L}_-^{(\vec{k})}+\alpha_0(\vec{k})\hat{L}_0^{(\vec{k})}]\right\}
%\end{equation}
%Here the operator $\hat{U}$ do not assume to be unitary.
%
In order to find  its complexity, we have to compute the lengths of all curves connecting $\hat{U}$ and identity in $\mathcal{U}$, and then find the minimal value of them. In $\mathcal{U}$, any curve starting from identity can be given by an $s$-dependent operator $\hat{O}(s)$ as
\begin{equation}\label{TandU1}
  \hat{O}(s)=\overleftarrow{P}\exp\left[\int_0^s\hat{T}(\tilde{s})\td \tilde{s} \right]\,,
\end{equation}
where
\begin{equation}\label{FGgenerator}
  \hat{T}(\tilde{s})=\int\td^{d-1}k[\alpha_+(\tilde{s},\vec{k})\hat{L}_+^{(\vec{k})}+\alpha_0(\tilde{s},\vec{k})\hat{L}_0^{(\vec{k})} +\alpha_-(\tilde{s},\vec{k})\hat{L}_-^{(\vec{k})}]\,.
\end{equation}
As $[\hat{T}(\tilde{s}_1),\hat{T}(\tilde{s}_2)]\neq0$ in general when $s_1\neq s_2$, the time-order operator cannot be neglected. Different choices of functions $\{\alpha_\pm(\tilde{s},\vec{k}), \alpha_0(\tilde{s},\vec{k})\}$ give different curves. We need this curve to end at $\hat{U}$ when $s=1$, i.e., $\hat{O}(1)=\hat{U}$.
Let us find the relationship between  $\{\gamma_\pm(\vec{k}), \gamma_0(\vec{k})\}$  defined in Eq.~\eqref{decomp1} and $\{\alpha_\pm(s,\vec{k}),\alpha_0(s,\vec{k})\}$ when we require that $\hat{O}(1)=\hat{U}$.

It is more convenient to consider the problem in the discrete momentum system. As the elements with different momentum in \eqref{TandU1} are commutative to each others, we can see that
\begin{equation}\label{TandU1b}
  \hat{O}(s):=\prod_{\vec{k}_i}\hat{O}_{\vec{k}_i}(s)\,,
\end{equation}
where
\begin{equation}\label{defUk}
  \hat{O}_{\vec{k}_i}(s):=\overleftarrow{P}\exp\left[\int_0^s\hat{T}_{\vec{k}_i}(\tilde{s})\td \tilde{s} \right]\,,
\end{equation}
\begin{equation}\label{FGgenerator}
  \hat{T}_{_{\vec{k}_i}}(s)=\alpha_+(\tilde{s},\vec{k}_i)\hat{L}_+^{(\vec{k}_i)}+\alpha_0(\tilde{s},\vec{k}_i)\hat{L}_0^{(\vec{k}_i)} +\alpha_-(\tilde{s},\vec{k}_i)\hat{L}_-^{(\vec{k}_i)}\,.
\end{equation}
The operator $\hat{O}_{\vec{k}_i}(s)$ has also a normal decomposition by three functions $b_{\vec{k}_i}(s), c_{\vec{k}_i}(s)$ and $d_{\vec{k}_i}(s)$
\begin{equation}\label{discdecomp111}
  \hat{O}_{\vec{k}_i}(s)=\exp[b_{\vec{k}_i}(s)\hat{L}_+^{(\vec{k}_i)}]\exp[c_{\vec{k}_i}(s)\hat{L}_0^{(\vec{k}_i)}]\exp[d_{\vec{k}_i}(s)\hat{L}_-^{(\vec{k}_i)}] \,.
\end{equation}

Differentiating  both  \eqref{defUk} and \eqref{discdecomp111}  with respective to $s$, we have
\begin{equation}\label{deriva111}
\begin{split}
  \hat{T}_{{\vec{k}_i}}\hat{O}_{\vec{k}_i}&=b'_{\vec{k}_i}\hat{L}_+^{(\vec{k}_i)}\exp[b_{\vec{k}_i}\hat{L}_+^{(\vec{k}_i)}]\exp[c_{\vec{k}_i}\hat{L}_0^{(\vec{k}_i)}]\exp[d_{\vec{k}_i}\hat{L}_-^{(\vec{k}_i)}]\\
  &+c_{\vec{k}_i}'\exp[b_{\vec{k}_i}\hat{L}_+^{(\vec{k}_i)}]\hat{L}_0^{(\vec{k}_i)}\exp[c_{\vec{k}_i}\hat{L}_0^{(\vec{k}_i)}]\exp[d_{\vec{k}_i}\hat{L}_-^{(\vec{k}_i)}]\\
  &+d_{\vec{k}_i}'\exp[b_{\vec{k}_i}\hat{L}_+^{(\vec{k}_i)}]\exp[c_{\vec{k}_i}\hat{L}_0^{(\vec{k}_i)}]\hat{L}_-^{(\vec{k}_i)}\exp[d_{\vec{k}_i}\hat{L}_-^{(\vec{k}_i)}]\,.
  \end{split}
\end{equation}
There is a very useful formula when we compute the right hand of \eqref{deriva111}. For any two operators $\hat{A},\hat{B}$, let us define  $[^{(0)}A,B]:=B$ and $[^{(n+1)}A,B]:=[A,[^{(n)}A,B]]$.  Then we can find
\begin{equation}\label{adjointf}
  e^{\hat{A}}\hat{B}e^{-\hat{A}}=\sum_{n=0}^{\infty}\frac1{n!}[^{(n)}\hat{A},\hat{B}].
\end{equation}
With this formula and the commutation relation~\eqref{su11Lie},  Eq.~\eqref{deriva111} becomes
\begin{eqnarray} %\label{deriva1}
%\begin{split}
  \hat{T}_{{\vec{k}_i}}(s)\hat{O}_{\vec{k}_i}&&=\left\{b'_{\vec{k}_i}\hat{L}_+^{(\vec{k}_i)}+c'_{\vec{k}_i}[\hat{L}_0^{(\vec{k}_i)}-b_{\vec{k}_i}\hat{L}_+^{(\vec{k}_i)}] +d_{\vec{k}_i}'e^{c_{\vec{k}_i}}[\hat{L}_-^{(\vec{k}_i)}-2b_{\vec{k}_i}\hat{L}_0^{(\vec{k}_i)}+b_{\vec{k}_i}^2\hat{L}_+^{(\vec{k}_i)}]\right\}\hat{O}_{\vec{k}_i} \nonumber \\
  &&=\left\{[b'_{\vec{k}_i}-c'_{\vec{k}_i}b_{\vec{k}_i}+b_{\vec{k}_i}^2d_{\vec{k}_i}'e^{c_{\vec{k}_i}}]\hat{L}_+^{(\vec{k}_i)} +[c'_{\vec{k}_i}-2d_{\vec{k}_i}'e^{c_{\vec{k}_i}}b_{\vec{k}_i}]\hat{L}_0^{(\vec{k}_i)}+d_{\vec{k}_i}'e^{c_{\vec{k}_i}}\hat{L}_-^{(\vec{k}_i)}\right\}\hat{O}_{\vec{k}_i} \nonumber \\
  &&=[\alpha_+(s,\vec{k}_i)\hat{L}_+^{(\vec{k}_i)}+\alpha_0(s,\vec{k}_i)\hat{L}_0^{(\vec{k}_i)} +\alpha_-(s,\vec{k}_i)\hat{L}_-^{(\vec{k}_i)}]\hat{O}_{\vec{k}_i}\,.  \nonumber
%  \end{split}
\end{eqnarray}
Thus we obtain the following differential equations
\begin{equation}\label{odebcd1}
\begin{split}
  &b'_{\vec{k}_i}(s)=\alpha_+(s,\vec{k}_i)+\alpha_0(s,\vec{k}_i) b_{\vec{k}_i}(s) + b^2_{\vec{k}_i}(s)\alpha_-(s,\vec{k}_i),\\
  &c'_{\vec{k}_i}(s)=\alpha_0(s,\vec{k}_i)+ 2b_{\vec{k}_i}(s)\alpha_-(s,\vec{k}_i),\\
  &d'_{\vec{k}_i}(s)=\alpha_-(s,\vec{k}_i)e^{-c_{\vec{k}_i}}\,.
  \end{split}
\end{equation}
They should satisfy the following boundary conditions:
\begin{equation}\label{reqonbcd}
\begin{split}
  &b_{\vec{k}_i}(0)=c_{\vec{k}_i}(0)=d_{\vec{k}_i}(0)=0,\\
  &b_{\vec{k}_i}(1)=\gamma_+(\vec{k}_i)\,,\quad c_{\vec{k}_i}(1)=\ln(\gamma_0(\vec{k}_i))\,, \quad d_{\vec{k}_i}(1)=\gamma_-(\vec{k}_i)\,.
  \end{split}
\end{equation}
The first line comes from the requirement $\hat{O}_{\vec{k}_i}(0)=I$  and the second line comes from $\hat{O}(1)=\hat{U}$ or $\hat{O}_{\vec{k}_i}(1)=\hat{U}_{\vec{k}_i}$, where $\hat{U}_{\vec{k}_i}$ is the discrete form of  Eq.~\eqref{decomp1}:
\begin{equation}
\hat{U}=\prod_{i}\hat{U}_{\vec{k}_i} \,,
\end{equation}
with
\begin{equation}\label{discdecomp1}
  \hat{U}_{\vec{k}_i}:=\exp[\gamma_+(\vec{k}_i)\hat{L}_+^{(\vec{k}_i)}]\exp[\ln\gamma_0(\vec{k}_i)\hat{L}_0^{(\vec{k}_i)}]\exp[\gamma_-(\vec{k}_i)\hat{L}_-^{(\vec{k}_i)}] \,.
\end{equation}

Based on the function form in Eq.~\eqref{defFp10},  the  complexity for a particular operator $\hat{U}$ defined in Eq.~\eqref{decomp1} can be obtained by following optimization problem
\begin{equation}\label{complCU2}
\begin{split}
  &\Sigma_{d-1}^{-1}\mathcal{C}(\hat{U}[\gamma_+(\vec{k}),\gamma_-(\vec{k}),\gamma_0(\vec{k})])\\
  =&\min\left\{\ell_0\int\td^{d-1}k\int_0^1\td s[\parallel\alpha_+(s,\vec{k})\parallel +\parallel\alpha_0(s,\vec{k})\parallel+\parallel\alpha_-(s,\vec{k})\parallel]\right\}
  \end{split}
\end{equation}
with the restrictions given by Eq.~\eqref{odebcd1} and Eq.~\eqref{reqonbcd}. As these restrictions are independent for different $\vec{k}$, we can further write Eq.~\eqref{complCU2} as
\begin{equation}\label{complCU3}
\begin{split}
  &\Sigma_{d-1}^{-1}\mathcal{C}(\hat{U}[\gamma_+(\vec{k}),\gamma_-(\vec{k}),\gamma_0(\vec{k})])\\
  =&\ell_0\int\td^{d-1}k\left[\min\int_0^1\td s\{\parallel\alpha_+(s,\vec{k})\parallel +\parallel\alpha_0(s,\vec{k})\parallel+\parallel\alpha_-(s,\vec{k})\parallel\}\right]
  \end{split}
\end{equation}
As the differential equations~\eqref{odebcd1} is highly nonlinear, for general values of $\gamma_+(\vec{k}),\gamma_-(\vec{k})$ and $\gamma_0(\vec{k})$, the optimization problem \eqref{complCU3} is not easy to solve.  However, it is possible to find the complexity presented in Eqs.~\eqref{FGTFDA2} and \eqref{goodset}. As the Eq.~\eqref{FGTFDA2} finds the minimal length in all the possible values of $\gamma_-(\vec{k})$ and $\gamma_0(\vec{k})$, the two of three functions $\alpha_\pm(s,\vec{k})$ and $\alpha_0(s,\vec{k})$ will be free. We can choose that $\alpha_-(s,\vec{k})$ and $\alpha_0(s,\vec{k})$ are free. Then Eqs.~\eqref{complCU3} becomes,
\begin{equation}\label{FGTFDA3}
\begin{split}
  \Sigma_{d-1}^{-1}\mathcal{C}(|\text{TFD}(t_L,t_R)\rangle,|0\rangle)&=\ell_0\int\td^{d-1}k\left[\min\int_0^1\td s\{\parallel\alpha_0(s,\vec{k})\parallel+\parallel\alpha_-(s,\vec{k})\parallel\right.\\
  &+\left.\parallel b'_{\vec{k}}-\alpha_0(s,\vec{k}) b_{\vec{k}}+b^2_{\vec{k}}\alpha_-(s,\vec{k})\parallel\}\right] \,,
  \end{split}
\end{equation}
for arbitrary functions  $\alpha^{(\vec{k})}_-, \alpha^{(\vec{k})}_0$ and $b_{\vec{k}_i}$ with $b_{\vec{k}_i}(0)=0$ and $b_{\vec{k}_i}(1)=\gamma_+(\vec{k})$. When $t=0$ in Eq.~\eqref{goodset}, the solution of \eqref{FGTFDA3} can be obtain by the following method. As $\gamma_+(\vec{k})\in\mathbb{R}$ and $0<\gamma_+(\vec{k})\leq1$, we can naturally expect that the solution of Eq.~\eqref{FGTFDA3} is given in the case $ b_{\vec{k}}(s)\in\mathbb{R}$ and $|b_{\vec{k}}(s)|\leq1$. Then one can see that
\begin{equation}\label{ineqs1}
\begin{split}
  &\int_0^1\td s\{\parallel\alpha_0^{(\vec{k})}(s)\parallel+\parallel\alpha_-^{(\vec{k})}(s)\parallel+\parallel b'_{\vec{k}}(s)-\alpha_0(s,\vec{k}) b_{\vec{k}}(s)-b^2_{\vec{k}}(s)\alpha_-(s,\vec{k})\parallel\}\\
  \geq&\int_0^1\td s\{\parallel b_{\vec{k}}(s)\alpha_0^{(\vec{k})}(s)\parallel+\parallel b^2_{\vec{k}}(s)\alpha_-^{(\vec{k})}(s)\parallel+\parallel b'_{\vec{k}}(s)-\alpha_0(s,\vec{k}) b_{\vec{k}}(s)-b^2_{\vec{k}}(s)\alpha_-(s,\vec{k})\parallel\}\\
  \geq&\int_0^1\td s\{\parallel b_{\vec{k}}(s)\alpha_0^{(\vec{k})}(s)+b^2_{\vec{k}}(s)\alpha_-^{(\vec{k})}(s)+ b'_{\vec{k}}(s)-\alpha_0(s,\vec{k}) b_{\vec{k}}(s)-b^2_{\vec{k}}(s)\alpha_-(s,\vec{k})\parallel\}\\
  =&\int_0^1\td s|b_{\vec{k}}'(s)|\geq|\int_0^1b_{\vec{k}}'(s)\td s |=|\gamma_+(\vec{k})|\,.
  \end{split} \nonumber
\end{equation}
The final equality can be satisfied only when $\alpha_0(s,\vec{k})=\alpha_-(s,\vec{k})=0$ and $b_{\vec{k}}'(s)<0$ or $b_{\vec{k}}'(s)>0$ for $\forall s\in(0,1)$. When $t_L+t_R\neq0$, we see Im$\gamma_+(\vec{k})\neq0$. In this case, we have to separate  every  variable into the real part and the imaginary part firstly. Then we use the definition in Eq.~\eqref{complexY} to convert $\parallel\cdot\parallel$ into the usual absolute symbol. After that, we can use the Euler-Lagrange equation to find the minimal value in Eq.~\eqref{FGTFDA3}. The result still shows that the minimal value can be reached if  $\alpha_0(s,\vec{k})=\alpha_-(s,\vec{k})=0$. Hence, we find that,
\begin{equation}\label{FGTFDA4}
  \Sigma_{d-1}^{-1}\mathcal{C}(|\text{TFD}(t_L,t_R)\rangle,|0\rangle)=\ell_0\int\td^{d-1}k\parallel\gamma_+(\vec{k})\parallel
\end{equation}
%

%==============================================================================
\bibliographystyle{JHEP}
\bibliography{QFTref}

\end{document}